\documentclass{article}

\usepackage{arxiv}
\usepackage{amsmath}
\usepackage[utf8]{inputenc} 
\usepackage[T1]{fontenc}    
\usepackage{hyperref}       
\usepackage{url}            
\usepackage{booktabs}       
\usepackage{amsfonts}       
\usepackage{nicefrac}       
\usepackage{microtype}      
\usepackage{lipsum}		
\usepackage{graphicx}
\usepackage{natbib}
\usepackage{doi}

\title{Sines, Transient, Noise Neural Modeling of Piano Notes}

\date{} 					

\author{ {\hspace{1mm}Riccardo Simionato} \\
	Department of Musicology\\
	University of Oslo\\
	Oslo, Norway \\
	\texttt{riccardo.simionato@imv.uio.no} \\
	\And
	{\hspace{1mm}Stefano Fasciani} \\
	Department of Musicology\\
	University of Oslo\\
	Oslo, Norway \\
	\texttt{stefano.fasciani@imv.uio.no} \\
}

\hypersetup{
pdftitle={A template for the arxiv style},
pdfsubject={q-bio.NC, q-bio.QM},
pdfauthor={David S.~Hippocampus, Elias D.~Striatum},
pdfkeywords={First keyword, Second keyword, More},
}

\begin{document}
\maketitle

\begin{abstract}
This article introduces a novel method for emulating piano sounds\sloppy. We propose to exploit the sines, transient, and noise decomposition to design a differentiable spectral modeling synthesizer replicating piano notes. Three sub-modules learn these components from piano recordings and generate the corresponding harmonic, transient, and noise signals. Splitting the emulation into three independently trainable models reduces the modeling tasks' complexity. The quasi-harmonic content is produced using a differentiable sinusoidal model guided by physics-derived formulas, whose parameters are automatically estimated from audio recordings. The noise sub-module uses a learnable time-varying filter, and the transients are generated using a deep convolutional network. From singular notes, we emulate the coupling between different keys in trichords with a convolutional-based network. Results show the model matches the partial distribution of the target while predicting the energy in the higher part of the spectrum presents more challenges. The energy distribution in the spectra of the transient and noise components is accurate overall. While the model is more computationally and memory efficient, perceptual tests reveal limitations in accurately modeling the attack phase of notes. Despite this, it generally achieves perceptual accuracy in emulating single notes and trichords.
\end{abstract}

\section{Introduction}\label{sec:intro}

Piano sound synthesis is a challenging problem and is still a topic of great interest. Historically, accurate piano models have been developed mainly using physical modeling techniques ~\citep{chabassier2014time}, which are required to discretize and solve partial differential equations. This approach leads to significant computational challenges. Recently, neural-based approaches, based on autoregressive architecture~\citep{hawthorne2018enabling} or Differentiable Digital Signal Processing (DDSP) framework ~\citep{engel2020ddsp}, have produced convincing audio signals. In particular, the DDSP framework, which involves integrating machine learning techniques into digital signal processors, has been mainly explored. Among traditional synthesis algorithms, DDSP has been integrated with wavetable\sloppy ~\citep{shan2022differentiable} and frequency modulation (FM) ~\citep{caspe2022ddx7} synthesis. DDSP allows data generation conditioned by MIDI information, specifically generating parameters used for digital signal algorithms. 
DDSP has also been used with Spectral Modelling Synthesis ~\citep{serra1990spectral} (SMS), which considers the sound signal as the sum of harmonic and noise parts. The harmonic content can be synthesized with a sum of sine waves whose parameters are directly extrapolated from the target. The noise is usually approximated with white noise processed with a time-varying filter. This method has been applied to ~\citep{kawamura2022differentiable} for a mixture of harmonic audio
signals, guitar ~\citep{wiggins2023differentiable, jonason2023ddsp}, and piano synthesis ~\citep{renault2022differentiable}. The method has been extended considering transients for the case of sound effects ~\citep{liu2023ddsp} and percussive sound synthesis ~\citep{shier2023differentiable}. In the first-mentioned work~\citep{liu2023ddsp}, the transients are synthesized using the discrete cosine transform (DCT) ~\citep{verma2000extending}, transforming an impulse signal in a sine wave. The authors synthesized the transient by learning the amplitudes and frequencies of its discrete cosine transforms and synthesized it using sinusoidal modeling. In the second work ~\citep{shier2023differentiable}, the transients are added to the signal using a Temporal convolutional neural network.
In the reviewed works, audio is mostly generated with a sampling frequency of $16$ kHz, except in ~\citep{liu2023ddsp} and \citep{jonason2023ddsp}, where the sampling rates are $22.5$ and $48$ kHz, respectively.

Artificial neural networks can learn complex system behaviors from data, and their application for acoustic modeling has been beneficial. Differentiable synthesizers aiming to synthesize acoustic instruments have already been proposed, but, as we have previously detailed, reviewed studies on guitar and piano neural synthesis employ models that learn from datasets filled with recorded songs and melodies. This approach increases the complexity of the tasks and can limit the network's ability to generalize, leading to challenges in scenarios not represented in the datasets. One example is the reproduction of isolated individual notes, where other concurrent sonic events might obscure the harmonic content of single notes in the dataset. Consequently, neural networks require a large amount of internal trainable parameters and training data to learn these complex patterns. 

The work we detail in this paper extends our prior proposal for modeling the harmonic contents of piano notes ~\citep{simionato2024physics}, and faces the problem from another perspective. We address the mentioned shortcoming by first emulating individual piano notes and then by incorporating the effects of key coupling, re-striking of played keys, and other scenarios that could occur during a real piano performance. Here, we combine the Differentiable Digital Signal Processing (DDSP) approach with the sines, transients, and noise (STN) decomposition ~\citep{verma2000extending} to synthesize the sound of individual piano notes, while a convolutional-based network is employed to emulate trichords. The generation of the notes is guided by physics knowledge, which allows designing a model with significantly fewer internal parameters. In addition, the model predicts a few audio samples at a time, allowing interactive low-latency implementations. The long-term objective of this research is to develop novel approaches for piano emulation that improve fidelity compared to existing techniques and are less computationally expensive than physical modeling while requiring less memory than sample-based methods.

Our approach includes three sub-modules, one for each component of the sound: harmonic, transient, and noise. The harmonic component corresponds to the partials during the quasi-stationary part of the note, which decays exponentially. The transient component, also referred to as percussive ~\citep{driedger2014extending}, refers to the initial, inharmonic, and percussive portion produced by the hammer strike, which decays rapidly. Lastly, the noise component is the residual from the decomposition process, which, in the case of piano notes, can be associated with background noise and friction sound arising from internal mechanisms, such as the movements of the hammer and key.

The harmonic component is modeled using differentiable spectral modeling to synthesize the sines waves composing the sound. The quasi-harmonic module, guided by physics formulas, predicts the inharmonic factor to produce the characteristic inharmonicity of the piano sound and the amplitude's envelope for each partial. The model also considers phantom partials, beatings, and double decay stages. The transients are modeled by generating a sine-based waveform that is transformed using the Inverse Discrete Cosine Transform (IDCT), producing the impulsive component of the sound. The vector of samples representing the waveform is computed entirely using a convolutional-based neural network. Finally, the noise component is synthesized filtering in the frequency domain of a Gaussian noise signal. The noise sub-model predicts the coefficients of the noise filter. Once the notes are computed, we emulate the coupling between different keys for trichords. A convolutional neural network processes the sum of individual note signals to emulate the sound of real chords. 

The approach synthesizes the notes separately, removing the need to have a large network learning how to synthesize all the notes using the same parameters. In this way, we need one model per key, similar to the sample-based methods, but we need to store a few parameters without storing the audio. Interpolation between different velocities, which indicate the speed and force with which a key is pressed to play a note, is also automatically modeled. This does not preclude the modeling of coupling or re-striking of keys, which is not allowed by sample-based methods, by adding an additional small network and, in turn, without adding significant complexity as in physical modeling methods.

The rest of the paper is organized as follows. Section ~\ref{sec:physics} describes the physics of the piano, focusing on aspects used to inform our model. Section ~\ref{sec:method} details our methodology, the architectures, the used datasets, and the experiments we have carried out. Section ~\ref{sec:results} summarizes and discusses the results we obtained, while Section ~\ref{sec:conclusion} concludes the paper with a summary of our findings.

\section{The Piano}
\label{sec:physics}

The piano is an instrument that produces sound by striking strings with hammers. When a key is pressed, a mechanism moves the corresponding hammer, usually covered with felt layers, to accelerate it to a few meters per second ($1$ to $6$~m/s) before release. One important aspect of piano strings is the tension modulation. This modulation occurs due to the interaction between the string's mechanical stretching and viscoelastic properties. It allows for shear deformation and changes in length during oscillation. These length changes cause the bridge to move sideways, which excites the string to vibrate in a different polarization. This vibration gains energy through coupling and induces the creation of longitudinal waves~\citep{etchenique2015coupling}.
Longitudinal motion ~\citep{nakamura1994characteristics, conklin1999generation} is also generated due to the string stretching during the collision with the hammer. When the hammer comes into contact with the string, it causes the string to elongate slightly from its initial length. This stretching creates longitudinal waves that propagate freely. The amplitude of these waves is relatively small compared to the transverse vibrations. The longitudinal vibration of piano strings greatly contributes to the distinctive character of low piano notes, and their properties and generation are not fully understood yet. They are named phantom partials and are generated by nonlinear mixing, and their frequencies are the sum or difference of transverse model frequencies. We can distinguish two types: even phantoms and odd phantoms. Even phantoms have double the frequency of a transverse mode, while odd ones have a frequency as the sum or difference of two transverse modes. Even and odd are pointed as the free and forced responses of the system ~\citep{bank2003modeling}. 

Another important phenomenon in piano strings happens when keys feature multiple strings~\citep{weinreich1977coupled}. Imperfections in the hammers cause the strings to vibrate with slightly varying amplitudes. In the case of two strings, they initially vibrate in phase, but each string loses energy faster than when it vibrates alone. As the amplitude of one string approaches zero, the bridge continues to be excited by the movement of the other string. As a result, the first string reabsorbs energy from the bridge, leading to a vibration of the opposite phase. This antiphase motion of the strings creates the piano's aftersound. Any discrepancies in tuning between the strings contribute to this effect, together with the bridge's motion, which can cause the vibration of one string to influence others. These phenomena affect the decay rates and produce the characteristic double decay.
Similarly, the same effects can be caused by different polarizations in the string. Specifically, since the vertical polarization is greater than that one parallel to the soundboard, more energy is transferred to the soundboard, resulting in a faster decay. When the vertical vibrations are small, the horizontal displacement becomes more significant, producing double decay. In addition, the horizontal vibration is usually detuned with respect to the vertical one ($\pm 0.1$ or $0.2$ Hz~\citep{tan2017piano}), creating beatings. 

String stiffness leads to another significant aspect characterizing the piano sound: the inharmonicity of its spectrum ~\citep{podlesak1988dispersion}. Specifically, the overtones in the spectra shift, so their frequencies do not have an integer multiple relationship. For this reason, they are called \textit{partials} instead of harmonics. There are slightly different formulations of this aspect, which lead to different definitions of the inharmonicity factor $B$; in this work, we consider the following formula derived from Timoshenko beam equations~\citep{chabassier2014time}:
\begin{equation}
    f_m = m F0 (1 + B m^2). \label{eq:B}
\end{equation}
where $B$ determines the degree of sound's inharmonicity, $m$ represents the partial number, and $F0$ is the fundamental frequency, which would be observed in the ideal case of a string without stiffness. Different derivations of $B$ exist, with variations often distinguished by the term $2\Big(1 - \frac{T_0}{E A}\Big)$, where $A$ denotes the cross-sectional area of the string, $T_0$ is its tension at rest, and $E$ represents the young modulus. For piano strings, $\frac{T_0}{E A}$ is much less than 1. Slightly different formulas for the inharmonicity factor are found in literature, and determining which one more closely matches a real piano remains an open challenge. However, the differences between these formulas are negligible~\citep{chabassier2014time}.  In addition, the precise formulation of $B$ is not a primary concern for this modeling approach, as the parameters are learned to closely replicate the particular sound of the piano in the given dataset.
Lastly, another important factor in the piano sound is the frequency-dependent damping due to energy losses while the string vibrates. One physically justified model is the Valette \& Cuesta (VC) damping model ~\citep{valette1988evolution}. In this model, the damping parameter is associated with the quality factor reported in Equation ~\ref{eq:VCM}:
\begin{equation}
    Q_m = \frac{\pi f_m}{\sigma_m} \label{eq:VCM}
\end{equation}
where $f_m = \omega/2\pi$ and $\sigma_m$ is the decay rate of the $m$-th partial.
\begin{equation}
    \frac{1}{Q_m} =  \frac{1}{Q_{m, air}} +  \frac{1}{Q_{m, vis}} +  \frac{1}{Q_{m, ther}} \label{eq:VCM2}
\end{equation}
$Q_{m, air}$, $Q_{m, vis}$, and $Q_{m, ther}$ are, respectively, the quality factors associated with losses due to air friction, viscoelasticity, and thermoelasticity of the string ~\citep{issanchou2017modal}. The three terms in Equation ~\ref{eq:VCM2} are described by Equation~\ref{eq:quality}:
\begin{align}
    \frac{1}{Q_{m, air}} &=  \frac{2 \pi \eta_{air} + 2 \pi d \sqrt{\pi \eta_{air} \rho_{air} f_m}}{2\pi \rho S f_m} \nonumber\\
    \frac{1}{Q_{m, vis}} &=  \frac{4 \pi^2 \rho S E I f_m^2}{T_0^2}\delta_{vis}\label{eq:quality}
\end{align}
while $Q_{m, ther}$ is constant. $\eta_{air}$ and  $\rho_{air}$ are the dynamic viscosity and density of air, $\rho$ is the string density, $\delta_{vis}$ is the viscoelastic loss angle, $d$ the diameter of the string, $S$ the cross-sectional area, and $I$ the area moment of inertia.

The soundboard damping is a linear phenomenon of modal nature, and the attenuation increases with the frequency of the modes ~\citep{ege2009table}. Therefore, the damping function is supposed to be increasing and have a sub-linear growth at infinity:
\begin{equation}\label{eq:soundboard}
    S_{loss}(f)  \leq C f + D
\end{equation}
with $C$ and $D$ $> 0$. 

Lastly, aspects of sound radiation are not considered here and are left as a separate and independent modeling challenge. For this purpose, ~\cite{renault2022differentiable} utilize a collection of room impulse responses, while recent advancements in neural approaches have also been explored ~\citep{mezza2024data}. 

\section{Methodology}
\label{sec:method}

The proposed modeling method presents three trainable sub-modules synthesizing different aspects of the piano sound. The respective quasi-harmonic, transient, and noise components extracted from real piano recordings are utilized to train the sub-modules separately. The sum of the signals generated by the three sub-modules represents the output of the model.
Each sub-module is fed at the input with a vector, including the fundamental frequency and the velocity of the note to be synthesized. For the sub-models synthesizing the quasi-harmonic part and the noise, the input vector also includes an integer time index informing the model of the time elapsed from the note's attack. The index expresses time in audio sampling periods, progressively incremented by the number of output audio samples predicted by each model's inference. Including this time index is beneficial for the model as it simplifies the task of tracking the envelopes in the target sound. Note that the model learns to generate the sound of individual piano notes with the same duration found in the dataset, which generally includes recordings of keys held in the struck position until the sound has fully decayed. Shorter durations can be obtained by integrating an envelope generator as an additional damping module and leveraging its release stage.
\begin{figure}[h!t]
\centerline{\includegraphics[scale=0.5]{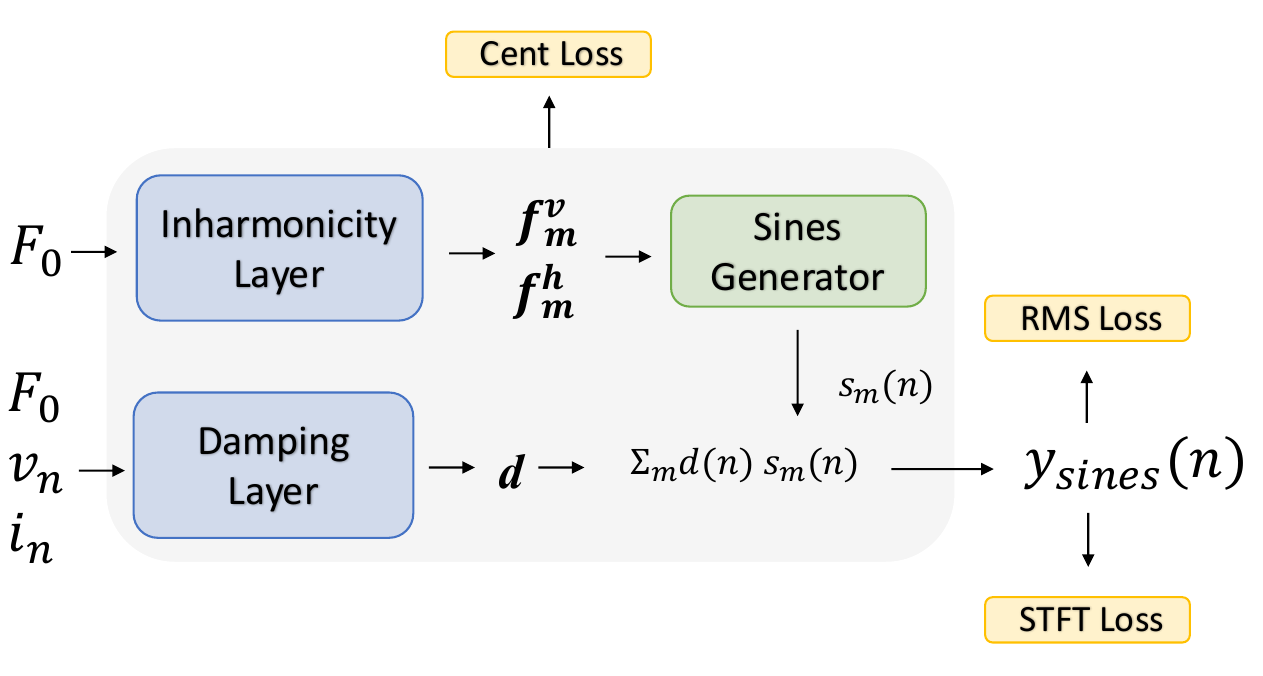}}
\caption{\label{fig:S}{\it The Quasi-Harmonic Model consists of $3$ layers: the Inharmonicity layer (a), the Damping layer (b), and the Sine Generator layer (c). The Inharmonicity layer computes the partial distribution for vertical and horizontal polarizations, the Damping layer predicts the partial amplitudes, while the Sine Generator layer computes and sums together all the sine components. 
}}
\end{figure}

\subsection{Quasi-Harmonic Model}\label{sec:harm}

\begin{figure*}[h!t]
\centerline{\includegraphics[scale=0.5]{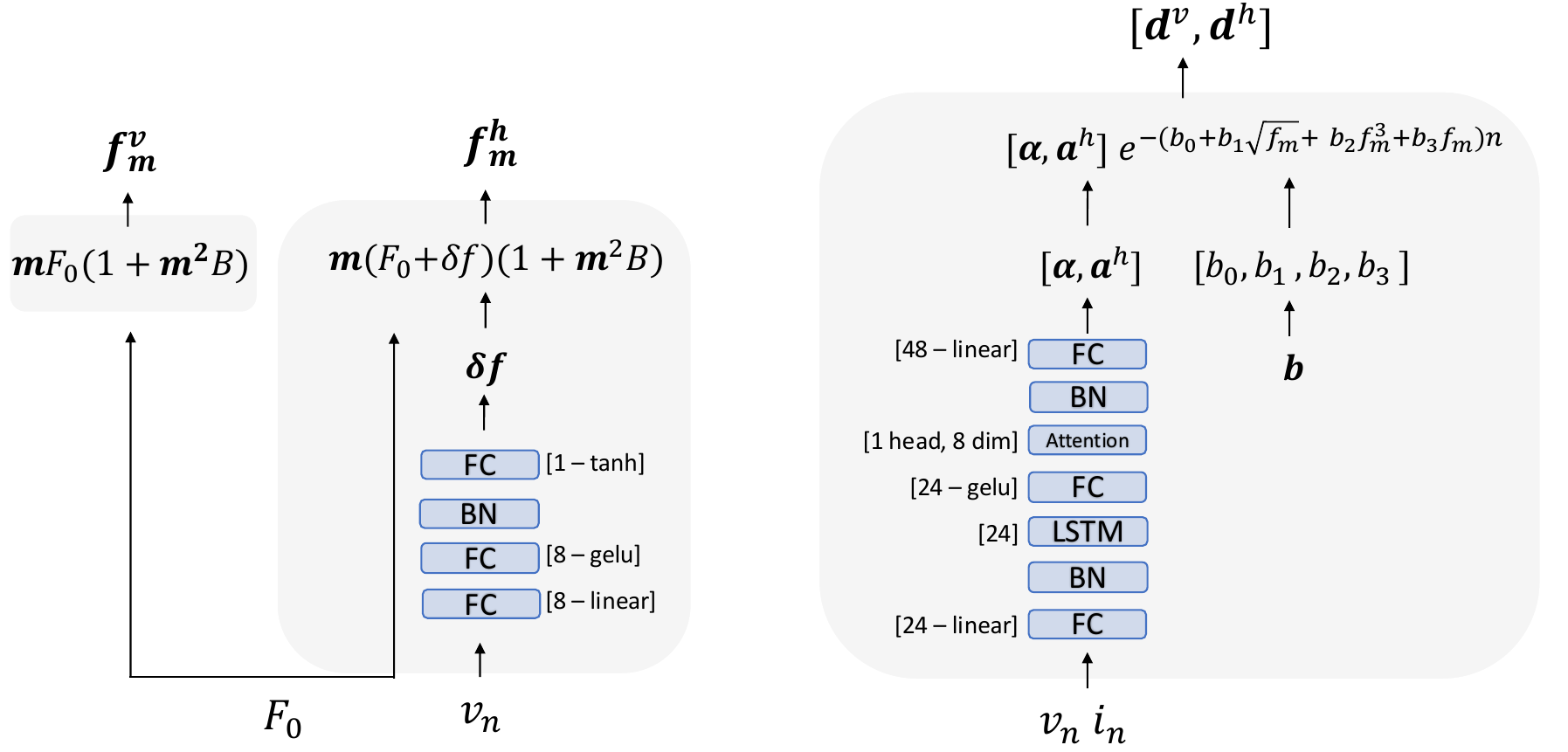}}
\caption{\label{fig:S1}{\it The layers composing the quasi-Harmonic Model. The Inharmonicity layer (left) takes the inharmonicity factor $B$, which is a learnable parameter, and computes the partial distribution for the vertical polarization $f^v_m$ from the input frequency. At the same time, the input velocity is fed to a feedforward network, which is used to detune the input frequency and predict the partial distribution for the horizontal polarization $f^h_m$. The Damping Layer (right) predicts the damping coefficients that govern the partial decaying for both polarizations. In this case, the input is the velocity and an index that indicates how many inference iterations have passed.}}
\end{figure*}

The architecture of the quasi-harmonic model is visible in Figure ~\ref{fig:S}, while the layers composing it is in Figure~\ref{fig:S1}. 
The quasi-harmonic model consists of $3$ layers: the Inharmonicity layer, the Damping layer, and the Sine Generator layer. Physical information is embedded in the model when generating the note's partials and envelope. The Sine Generator layer takes the partial frequency values and generates the corresponding sine waves. In the quasi-harmonic model, the learning is pursued using the multi-resolution Short-time Fourier Transform (STFT) loss and the Mean-Absolute-Error (MAE) of the Root-Mean-Square (RMS) energy computed per iteration, as similarly in ~\citep{simionato2024physics}. The multi-resolution STFT loss is the average of STFT losses computed with window sizes of [$256$, $512$, $1024$, $2048$, $4096$] and $25\%$ overlap. By doing so, the smallest frequency resolution is $93.75$ Hz. The loss is normalized by the norm of the target spectrogram, allowing for greater weight to be assigned to the note's release phase, even when it has a small amplitude. The losses compare the predicted quasi-harmonic content against the quasi-harmonic one separated from the real recording as explained in Section ~\ref{sec:data}. 

\subsubsection{Inharmonicity Layer}

The Inharmonicity layer adjusts the generation of partials based on the learnable inharmonicity factor $B$, using Equation ~\ref{eq:B}.
%
%
This approach provides an approximation to the inharmonic characteristics of the note. The tuning employs the Cent loss ~\citep{simionato2024physics}, which takes into account the first $6$ partials. The actual fundamental frequency $F0$ is estimated from the piano recordings in the dataset by using a peak estimation algorithm to detect the frequencies of the partials, as detailed in Section \ref{sec:data}.

\subsubsection{Damping Layer}

The Damping layer is based on the VC damping model. Partials exponentially decay over time as described by Equation ~\ref{eq:allsines}:
\begin{equation}
    y_{sines}(n) = \sum_m \alpha_m e^{-\sigma_m n} s_m(n) \label{eq:allsines}
\end{equation}
where $y_{sines}(n)$ is the quasi-harmonic content of the signal and $s_{m}$ the $m$-th partial, $\alpha_m$ is the initial amplitudes, and $\sigma_m$ it the decay rate of the $m$-th partial. Deriving from the information detailed in Section~\ref{sec:physics}, we can write:
\begin{align}
    \sigma_m = \pi f_m \big( \frac{1}{Q_{m, air}} +  \frac{1}{Q_{m, vis}} + \frac{1}{Q_{m, ther}}  \big) \label{eq:sigma}
\end{align}
where $Q_{m, air}$, $Q_{m, vis}$ are frequency-dependent and $Q_{m, ther}$ is constant. To integrate this information into the model, we simplify the expressions as follows:
\begin{equation}
         \frac{1}{Q_{m, air}} = \frac{b_0}{f_m}  + \frac{b_1}{\sqrt{f_m}}, \label{eq:q1}
\end{equation}
where $b_0 = \frac{2 \pi \eta_{air}}{2\pi \rho S}$ and  $b_1 = \frac{2 \pi d \sqrt{\pi \eta_{air} \rho_{air}}}{2\pi \rho S}$,
\begin{equation}
         \frac{1}{Q_{m, vis}} = b_2 f_m^2 \label{eq:q2}
\end{equation}
where $b_2 = \frac{4 \pi^2 \rho S E I}{T_0^2}\delta_{vis}$, and finally
\begin{equation}
     \frac{1}{Q_{m, ther}} = b_3, \label{eq:q3}
\end{equation}
where $b_3$ represent a constant which can be measured experimentally ~\citep{tan2017piano}. Substituting Equations ~\ref{eq:q1}, ~\ref{eq:q2}, ~\ref{eq:q3} in ~\ref{eq:sigma}, we obtain the following expression for the decay envelope:
\begin{align}
    \alpha_m e^{-\sigma_m} = \alpha_m e^{-\pi  (b_0 + b_1 \sqrt{f_m} + b_2 f_m^3 + b_3 f_m)n} \label{eq:damping}
\end{align}
where $\alpha_m$ is predicted by the Damping layer and $b_0$, $b_1$, $b_2$, and $b_3$ are tunable parameters. The $\alpha_m$ coefficients are predicted by a network fed with the velocity $vel$, and the index $i$, which indicates how many sampling periods have passed since the key was pressed. The network presents a linear Fully-Connected (FC) layer, a Long Short-term Memory (LSTM) layer, another FC layer featuring the Gaussian Error Linear Unit (GELU) ~\citep{hendrycks2023gaussian}, all having with $24$ units, an attention layer computing the attention across the partials dimensions, and an output layer having a number of units equal to the desired number of harmonics $H$. Batch normalization layers are placed before the LSTM and before the attention layers. From the prediction, the absolute value is clipped in case they are greater than $1$.
The modulus operation is applied to the learnable vector $b$, which is then split into the individual coefficients $b_0$, $b_1$, $b_2$, and $b_3$. These $b$ coefficients are unique to each note and are determined solely by the string's parameters, independent of the index $i$. By utilizing $b_0$, $b_1$, $b_2$, and $b_3$, we eliminate the need to include soundboard losses in the model, as they are implicitly accounted for by $b_0$ and $b_3$. When incorporating soundboard damping from Equation~\ref{eq:soundboard}, the bias and the term that depends linearly on frequency are modified to $b_0 + B$ and $b_3 + A$, respectively. However, for simplicity, we retain the notation of $b_0$ and $b_3$.

The actual decay rate for each partial is generated using Equation ~\ref{eq:damping}, and the corresponding sine waves are synthesized as described in Equation ~\ref{eq:allpartials}:
\begin{equation}
     y_{sines}(n) = \sum_{m=1}^{H} \alpha_m e^{-\sigma_m n} sin (2 \pi f_m n). \label{eq:allpartials}
\end{equation}
The model can generate amplitude values for audio segments of any length without architectural constraints. Using lower sample numbers leads to reduced latency, longer training times, and increased computational costs per sample during inference. However, the accuracy of the results tends to improve, as this approach avoids additional artifacts by allowing the network to update the damping coefficients more frequently. The results presented in this paper pertain to the extreme case where the model generates a single sample per inference iteration.

\subsection{Beatings and Double Decay}

As mentioned in Section ~\ref{sec:physics}, piano strings present vertical and horizontal vibrations. The vibrations have different decay rates. The string exchanges energy with the bridge. The vertical vibration energy dissipates quickly, making up the initial fast decay. As the vertical displacement reduces, horizontal displacement becomes more
dominant and exhibits the second slower decay. This created the so-called double decay, while the detuning between the two polarizations creates beatings. 
Similarly, the same effect can be created when the hammer strikes multiple strings. Strings in the same key cannot be perfectly in tune or slightly differently excited. The detuning creates out-of-phase vibrations that cancel each other out and reduce the energy exchange with the bridge. As before, it creates two different decaying rates and beatings.

This feature is integrated into the model, generating an additional set of partials. To account for the second polarization, the model employs a feedforward network consisting of two linear hidden layers with $8$ units each. The first layer operates without activation functions, while the second utilizes the GELU activation function. Batch normalization is placed before the output layer with a single unit and the hyperbolic tangent as the activation function. The network predicts the detuning factor $\delta f$ given the velocity $v_n$. The detuning factor is added to $F0$, and the result is used to compute the second set of partials for the horizontal polarization $f^h_m$, again using Equation ~\ref{eq:B}. The damping coefficients are generated using the same Damping layer. Specifically, from the predicted set of coefficients $\alpha_m$, two sets of coefficients, $\alpha^v_m$ and $\alpha^h_m$ are extracted. This approach allows the model to create two distinct decay shapes from the same network, which also determines the energy coupling between these sets of partials.

\subsection{Phantom Partials}

The model incorporates longitudinal displacements described in Section~\ref{sec:physics}. Instead of employing a machine-learning approach, phantom partials are computed directly from the predicted transverse partials and their associated coefficients. Specifically, even phantom partials are obtained by doubling the frequencies of the two transverse polarizations and squaring their corresponding coefficients, as described by Equation ~\ref{eq:l}:
\begin{equation}
     y^{l}_{even}(n) = \sum_{m=1}^{H} \alpha^2_m e^{-2\sigma_m n} sin (2 \pi 2 f_m n)\label{eq:l}
\end{equation}
Note that doubling the partials leads to doubling the decay rates $\sigma_m$.
The odd phantom partials are instead given by the sum and difference of consecutive transverse partials and the multiplication of the related coefficients as in ~\citep{bank2005generation}, and reported in Equations ~\ref{eq:lf} and ~\ref{eq:lcoeff}:
\begin{align}
     f^{l}_{k} &= f_n \pm f_{m} \label{eq:lf}\\
     \alpha^{l}_{k} &= \alpha_n  \alpha_{m} \label{eq:lcoeff}
\end{align} 
where $|n-m|=1$, and $k$ is the longitudinal mode index. 
Therefore, we obtain $5$ new partials sets: two sets related to even phantom partials and three sets related to odd phantom partials, both originated from the two sets of transverse partials. Lastly, since $f_k^l \geq 10 \cdot f_n$ ~\citep{chabassier2014time}, we added only the partials satisfying the criterion.

\subsection{Transient Model}\label{sec:t}

\begin{figure}[h!t]
\centerline{\includegraphics[scale=0.5]{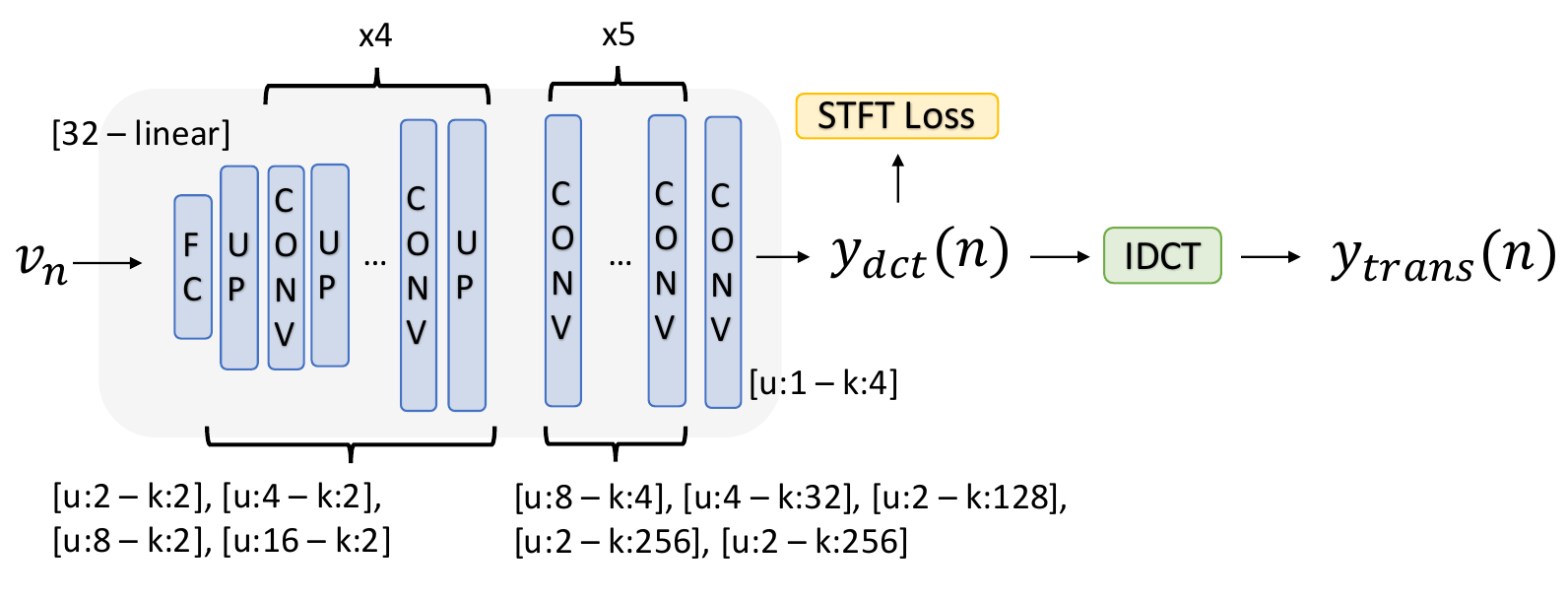}}
\caption{\label{fig:T}{\it The transient model takes the velocity of the note as input and, using a stack of upsampling and convolutional layers, generates the waveform. The resulting waveform is inversely discrete cosine transformed to compute $y_{trans}$.}}
\end{figure}

The architecture of the Transient model is illustrated in Figure~\ref{fig:T}. We exploit the discrete cosine domain to synthesize the transients. In particular, we designed a deep convolutional oscillator to compute a sines-based waveform that will be converted using an inverse Discrete Cosine Transform (IDCT). Similarly to \citep{krekovic2022deep}, we designed a generative network based on a stack of upsampling and convolutional layers to generate the waveform from the frequency and velocity information. The model outputs an array of $1300$ samples representing the DCT waveform that is converted into the transient using the Equation ~\ref{eq:trans}:
\begin{equation}
      y_{trans}(n) = IDCT(D(n)).\label{eq:trans}
\end{equation}
$D(n)$ is a sum of sines waves.

The multi-resolution STFT loss function compares the output vector to the actual DCT signal from the dataset, using [$32$, $64$, $128$, $256$] as window lengths. The input to the network is the velocity $v_n$. 
\subsection{Noise Model}

\begin{figure}[ht]
\centerline{\includegraphics[scale=0.5]{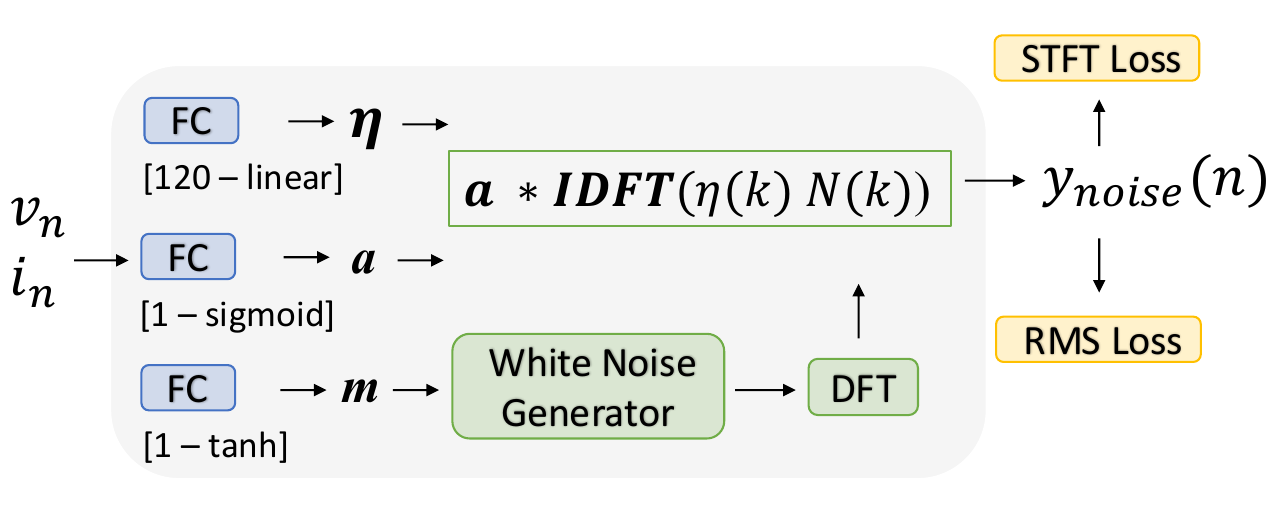}}
\caption{\label{fig:N}{\it The noise is modeled generating noise filter magnitudes $\boldsymbol{\eta}$ that is convolved in the frequency domain with a generated white noise. The input vector for all the layers consists of the velocity $v_n$, and time index $i_n$.}}
\end{figure}

The architecture of the Noise model is shown in Figure~\ref{fig:N}. The noise is modeled generating noise filter magnitudes $\eta(k)$ that filter a white noise signal ~\citep{engel2020ddsp}, as described by Equation ~\ref{eq:noise}:
\begin{equation}
     y_{noise}(n)= \boldsymbol{a} \cdot IDFT (\eta (k) N(k)) \label{eq:noise}
\end{equation}
where $N(k)$ is the Discrete Fourier Transform (DFT) transform of the noise and $y_{noise}(n)$ is the filtered output. A linear FC layer generates the filter magnitudes. The predicted and the target are compared using the multi-resolution STFT loss function, computed using window sizes of [$32, 64, 128, 256, 512$], and the MAE of the RMS energy.

To help the prediction, we also anchor the mean of the white noise signal to be generated. An FC layer with $1$ unit and hyperbolic tangent activation predicts the mean of the white noise to be generated, and another FC layer with $1$ unit and sigmoid activation computes the amplitudes of the noise signal. In this case, the MAE of the RMS is used as a loss function. The inputs of all the cases are a vector containing the velocity $v_n$ and index $i_n$.

\subsection{Trichords}\label{sec:board}

\begin{figure}[ht]
\centerline{\includegraphics[scale=0.5]{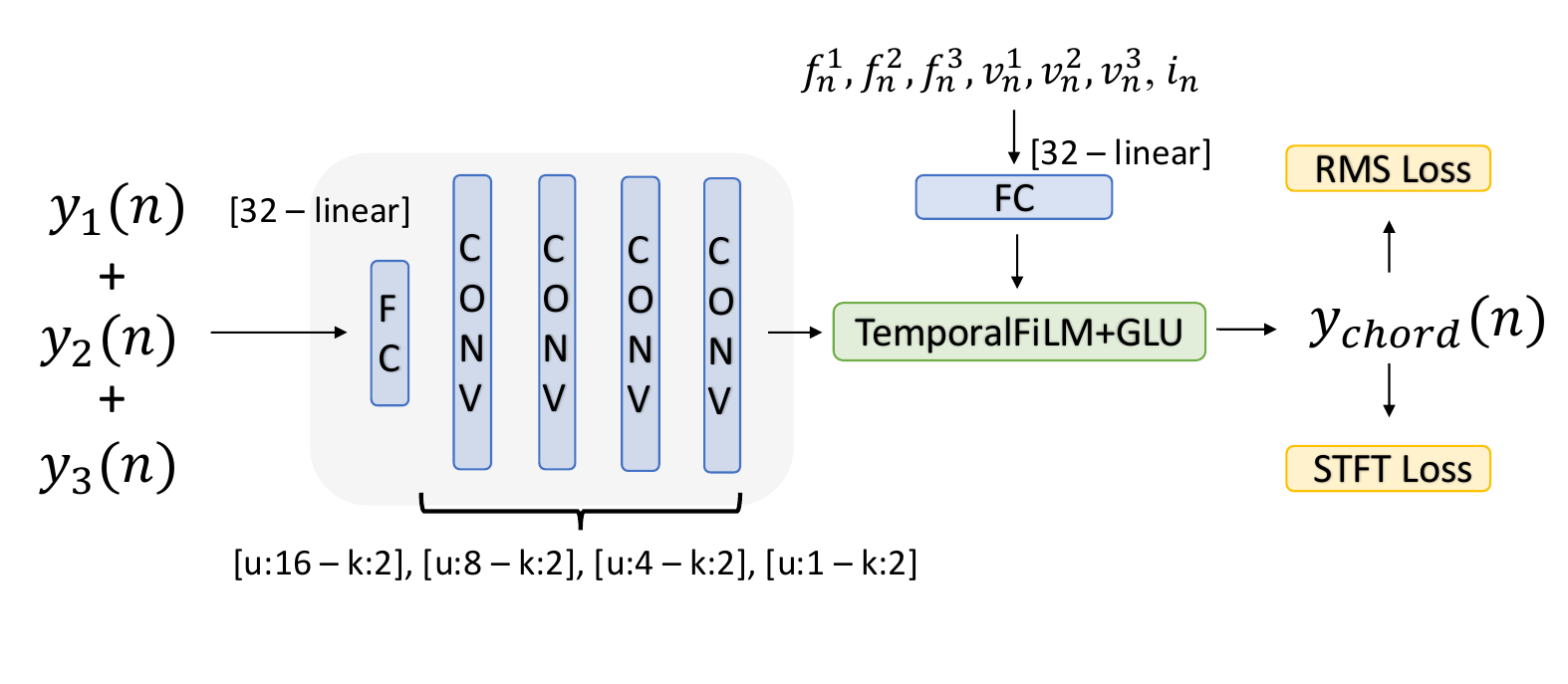}}
\caption{\label{fig:chords}{\it The coupling among different keys is modeled by a convolutional neural network and temporal FiLM method combined with the GLU to condition the network based on the keys playing. The input vector is the sum of the three separate note sounds, while their frequencies, velocities, and time index compose the conditioning vector.}}
\end{figure}

The coupling among different keys that happens during chords is modeled by a convolutional neural network and temporal Feature-wise Linear Modulation (FiLM) method ~\citep{birnbaum2019temporal} followed by Gated Linear Unit ~\citep{dauphin2017language} as in ~\citep{simionato2024conditioning}, which conditions the network based on the playing keys. The input vector consists of the sum of the signals of the three generated notes, while the frequencies and velocities corresponding to the three notes, together with the time index $i_n$, are concatenated to create the conditioning vector. A linear, fully connected layer with $32$ units processes the conditioning vector before being fed into the FiLM layer. The target is the real recorded trichord. For training the network, we use the MAE of the RMS energy and the multi-resolution STFT detailed in Section ~\ref{sec:harm}. In this case, the window sizes for the multi-resolution STFT loss are $256, 512$, and $1024$. Figure~\ref{fig:chords} shows the architecture. 

\subsection{Dataset}\label{sec:data}

For this study, we use the BiVib dataset\footnote{\url{https://zenodo.org/records/2573232}}, which includes piano sounds recorded from electro-mechanically actuated pianos controllable via MIDI messages. We selected the piano recordings related to the upright-closed and grand-open collection. For both collections, we also narrowed the recordings utilized to train the models to two octaves (C3 to B4), and we considered velocities between $45$ and $127$. Each recording in the dataset is associated not with a specific velocity value but rather with a velocity range, likely because small variations in velocity produce negligible changes in sound. For our experiments, we have associated each recording with the lower bound of its respective velocity range. This resulted in $168$ note recordings with [$45, 56, 67, 78, 89, 100, 111$] as velocity values and $24$ different notes, which are $10$ seconds long. Audio is downsampled to $24$ kHz, the rate at which our model generates audio samples. The first $6$ partials were extrapolated from the audio recordings using a manual local peak estimation. These partials are used to estimate the inharmonicity factor $B$, computed from Equation~\ref{eq:B} and exploiting the ratio between $f_m$ and $f_j$, as reported in Equation ~\ref{eq:Bestimation}:
\begin{equation}
\frac{f_m}{f_j} = \frac{m (1+B m^2)}{j (1+Bj^2)},
\label{eq:Bestimation}
\end{equation}
where $m$ and $j$ are two different partial numbers. This equation leads to the expression to compute $B$ and showed in Equation ~\ref{eq:Bestimation2}:
\begin{equation}
B = \frac{\frac{f_m}{f_j} j - m}{m^3 - \frac{f_m}{f_j} j^3}.
\label{eq:Bestimation2}
\end{equation}
We use all combinations derived from the $6$ extrapolated partials to estimate $B$. We obtain $30$ estimations of $B$ for each note per dataset recording. This process is iterated across all the $7$ velocities per note, obtaining $210$ values. We then take the mean of these estimations as the initial value for $B$ in our experiments. While this method may be prone to inaccuracies stemming from peak estimation errors and the simplifications made in deriving Equation~\ref{eq:B}, it provides a starting point that situates the model parameters near realistic values. The computed $B$ values are subsequently used to calculate $F0$, which serves as an input representing the frequency of the key under the assumption of no string stiffness.

The transient and harmonic content of the notes were constructed using the Harmonic Percussive Source Separation (HPSS) method ~\citep{driedger2014extending} with a margin of $8$. These separated components are considered the target of each sub-model. Once trained, the model will generate each component to be added to each other and reconstruct the original sound. The noise components were computed by subtracting the transient and harmonic spectrograms from the original note spectrogram. The DCT was also calculated from the transient component. 

For the chords, we extended the dataset used in ~\citep{simionato2024physics} by including not only singular notes but also minor triads in the C3-B4 range, played at velocities of [$60$, $70$, $80$, $90$, $100$, $110$, $120$]. The input-output pair, in this case, is the sum of separate notes and the actual trichord recording. 

\subsection{Experiments Setup}

The models are trained using the Adam ~\citep{kingma2014adam} optimizer with a gradient norm scaling of 1 ~\citep{pascanu2013difficulty}. The training was stopped earlier in case of no reduction of validation loss for $50$ epochs, and the learning rate was reduced by $25\%$ if there were no improvements after $1$ epoch. The initial learning rate was $3 \cdot 10^{-4}$. The test losses are computed using the model’s weights, which minimize the validation loss throughout the training epochs. The raw audio is split into segments of a $1$ sample and processed in batch sizes of $131072$ for the case of the single notes and $24000$ for the chords before updating the weights. In addition, the internal states of the LSTM layer are initialized when a different note needs to be generated. The training of the quasi-harmonic module is divided into two stages. Initially, only the inharmonic factor is tuned to determine the position of the partials, and in the second stage, the damping layer is considered.

The datasets considering single notes are partitioned by key, with the middle velocity of $78$ designated as the test set. On the other hand, in the trichords scenario, the dataset is split into $90\%$ for the training set and $10\%$ for the test set. Audio examples, source code, and trained models described in this paper are available online~\footnote{\url{https://github.com/RiccardoVib/STN_Neural}}.

\subsection{Perceptual Evaluation}

Perceptual evaluation is based on Multiple Stimuli test with Hidden Reference and Anchor (MUSHRA)  ~\citep{bs20151534}, which is utilized to evaluate single notes and trichords generated by the proposed model. Usually, the MUSHRA test involves multiple trials. In each trial, the listener is requested to rate a set of signals on a scale $0$ to $100$ against a given reference signal. Each $20$-point band on the scale is labeled with descriptors: excellent, good, fair, poor, and bad. Each trial requires participants to rate a set of signals, which includes the reference itself, known as the hidden reference, as well as one or more anchors. The hidden reference is used to identify clear and obvious errors. Indeed, data from listeners who rate the hidden reference lower than $90$ for more than $15\%$ of the trials are excluded. Anchors are intentionally impaired versions of the reference that participants should readily identify as such. They provide a context for the lower end of the quality scale. The standard recommendation specifies that the low-range and mid-range anchors should be low-pass versions of the reference at cutoff frequencies of $7$ kHz and $3.5$ kHz, respectively. This setting is appropriate when the reference signal contains frequency components with significant energy across the entire audible spectrum. If the reference lacks such high-frequency content, the cutoff frequency can be adjusted downward to ensure the anchor is a noticeably impaired version of the reference.

The test consisted of $9$ trials: in the first $6$, participants rated single notes generated by the models trained with datasets from both an upright and a grand piano, with $3$ trials dedicated to each. This was done to evaluate whether the model had successfully learned the characteristics of the specific target piano. In the last $3$ trials, participants rated trichords to assess the model’s ability to emulate a realistic-sounding trichord. The models are trained on a dataset with notes ranging from $C3$ to $B4$; therefore, both the target references and the predicted signals that are rated in the tests fall within this pitch range. Within this range, piano sounds typically have low energy in the higher portion of the spectrum, and low-passed anchors at the recommended cutoff frequencies would not result in a noticeably impaired version of the reference. For this reason, we have opted to use a single anchor, which is a low-passed version of the reference with a cutoff frequency of $1$ kHz.

To assess the statistical significance of the MUSRA test results, we employ the Friedman test and the Cliff’s delta. The Friedman test evaluates the consistency of rankings assigned to the MUSHRA scores across different trials. Scores from each trial are ranked in ascending order; the lowest score receives a rank of $1$, while the highest score is given a rank equal to the number of conditions evaluated in the trial. These ranks are used to calculate a test statistic, from which $p$-values are derived to assess statistical significance. Furthermore, Cliff’s delta is computed and measures the effect size $|d|$, which is considered small for $0.11 \leq |d| < 0.28$, medium for $0.28 \leq |d| < 0.43$, and large for $|d| \geq 0.43$. 

MUSHRA tests are carried out in a controlled, quiet room, with participants using professional over-ear headphones, the Beyerdynamic DT-990 Pro, connected to the output of a MOTU M4 audio interface. Participants are provided with a browser-based Graphical User Interface (GUI) ~\citep{schoeffler2018webmushra} to facilitate the process of listening, comparing, and rating the audio signals. The $9$ trials are conducted sequentially, in a fixed order, without interruption, and the total duration of the test, excluding the introduction and setup, is approximately $10$ minutes.

\subsection{Comparative Evaluations}

To further evaluate the approach, we employed the computational method proposed in \cite{simionato2023computational}. We used the same audio descriptors in the proposal but are considering recording at $24000$ Hz. In particular, the method involves the Linear Discriminant Analysis (LDA) projection to maximize the separability between the types of pianos, acoustic, sample-based models, and physics-based models while minimizing within-type variance. LDA projects the feature vectors into a lower-dimensional space with dimensionality or a number of components equal to the number of
classes minus one. A large set of audio descriptors are used to compute the feature vectors. For this paper, we considered two cases, the single note and trichord stimuli. For each type of stimulus, each feature vector is associated with a label representing the type of piano, acoustic or synthetic Pulse Code Modulation (PCM)-based or synthetic physics-based. For the single-note case, each vector is a high-dimensional representation of how the generated sound changes when increasing the velocity of the stimulus, while for the trichord case, each vector represents the difference between the sum of three notes and the actual trichord. For the single note case, we compute the dataset by creating a recording of sample-based and physical-based digital pianos, using the same velocities and the length of the dataset used for training our models. On the opposite, since the dataset used for training the trichord models uses the same sounds collected for the comparative study presented in \cite{simionato2023computational}, we downsampled the original example to $24000$ Hz.

\section{Results}
\label{sec:results}

\begin{figure*}[ht!]
\centering
\includegraphics[scale=0.55]{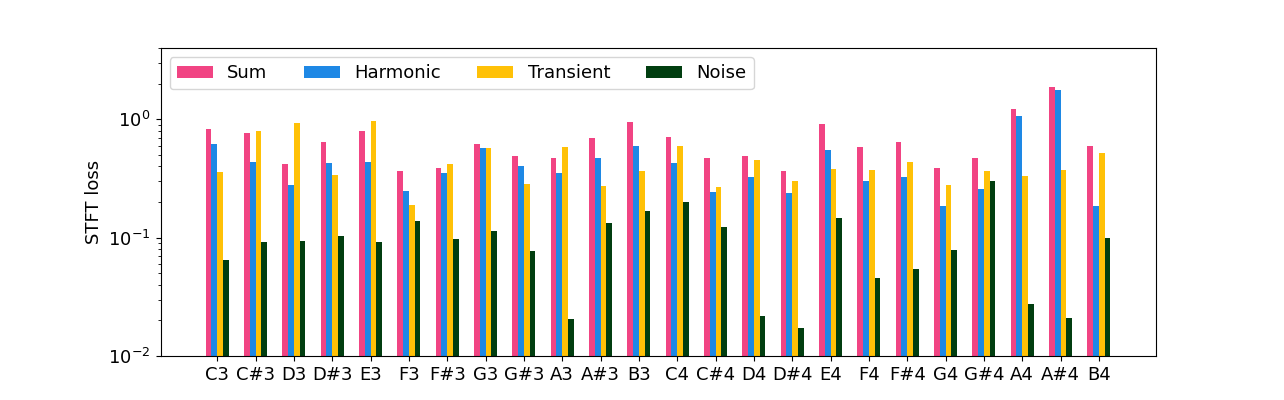}\\
\includegraphics[scale=0.55]{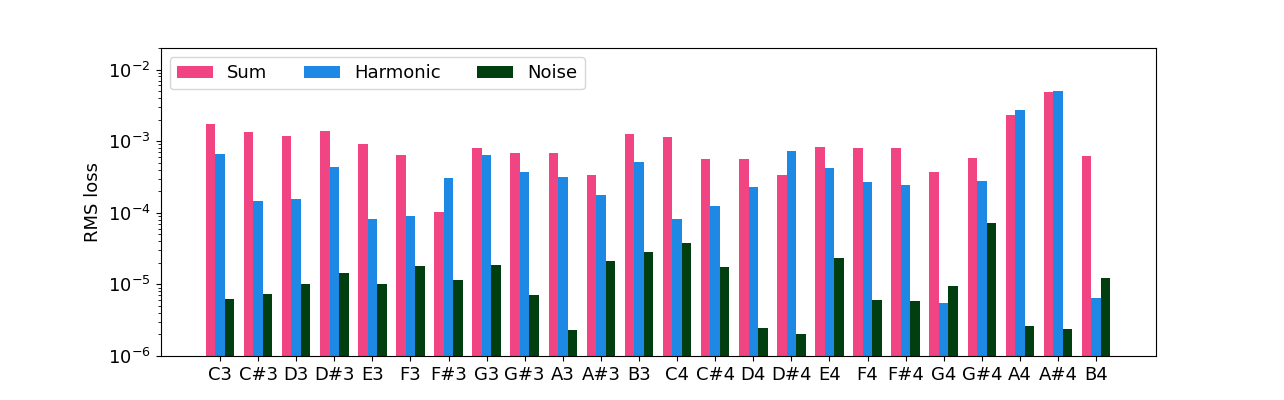}
\caption{\label{fig:up}{\it STFT (top) and RMS (bottom) losses of the test set across all the keys and considering the upright piano. STFT losses are normalized by the norm of the target spectrogram, as in the training process, and the window resolutions employed are identical to those used for the STFT of the quasi-harmonic component. The y-axis is on a logarithmic scale. Magenta bars refer to the sum of all components; blue bars refer to the quasi-harmonic component; the yellow bars refer to the transient component, and the dark green bars refer to the noise component.}}
\end{figure*}

Figure ~\ref{fig:up} shows the RMS and STFT losses for models of each key, also detailing losses for the quasi-harmonic, transient, and noise sub-models. The bar plot presents the STFT losses, normalized by the target spectrogram norm, as used for training the quasi-harmonic sub-module. The window resolutions utilized here are the same STFT resolutions used for the quasi-harmonic component. 
Generally, the model presents good accuracy considering all component-related losses. The models' accuracy is also consistent in most of the keys across the octaves, except for $A4$ and $A\#4$ keys, which appear to be more challenging for accurate modeling. We have also computed the same loss that was used to train the different sub-modules, comparing the sum of the three components to the actual piano recordings in the dataset. Here, inaccuracies from the decomposition process further contribute to increasing the total error, thereby having a negative impact. Generally, the loss computed on the sum is close to that of the quasi-harmonic sub-module because the transient signal is very short, and the noise signal has a very low energy. Specifically, the transient signal is non-zero only for a relatively short duration—1,300 samples out of the total length of the predicted notes, which is $240,000$ samples—while the quasi-harmonic signal is typically $22$ dB greater in magnitude than the noise signal. 

\begin{figure*}[ht!]
\centering
\includegraphics[scale=0.55]{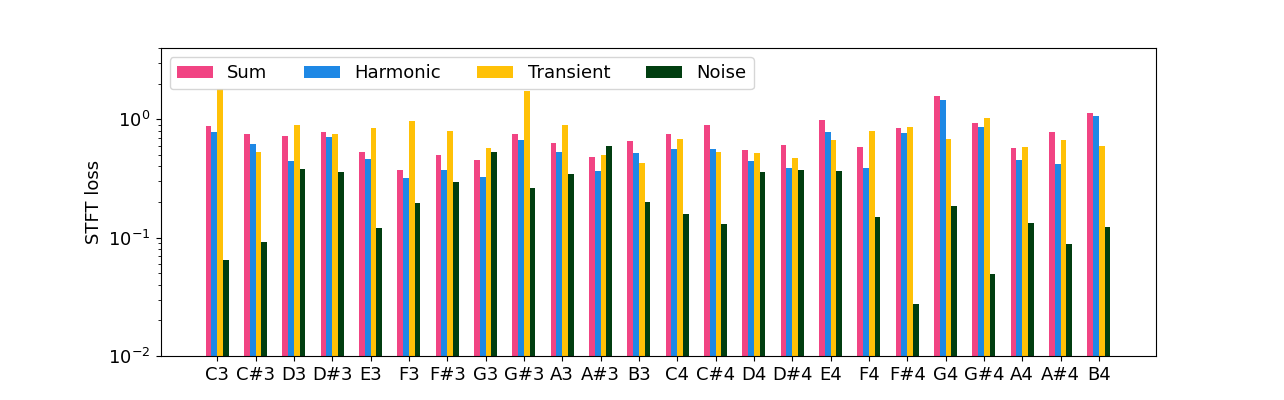}\\
\includegraphics[scale=0.55]{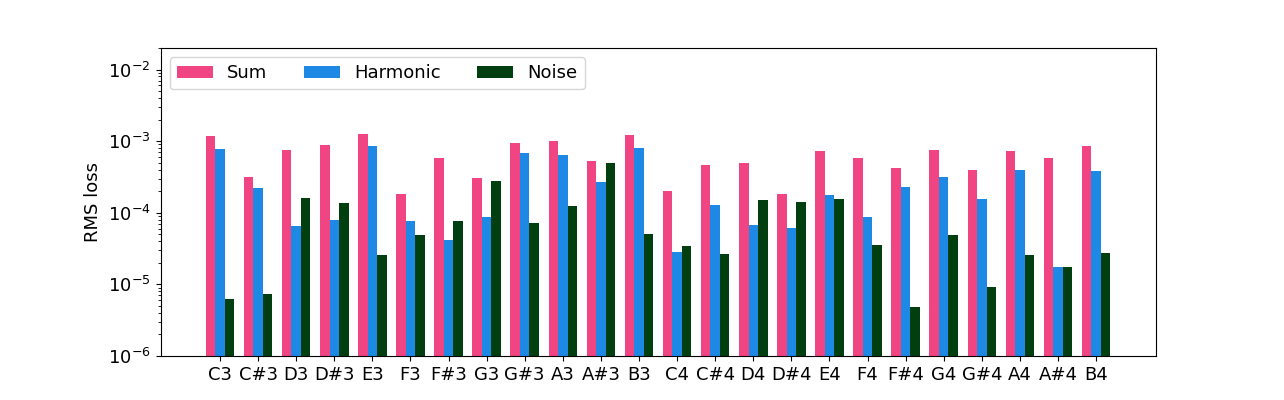}
\caption{\label{fig:grand}{\it STFT (top) and RMS (bottom) losses of the test set across all the keys and considering the upright piano. STFT losses are normalized by the norm of the target spectrogram, as in the training process, and the window resolutions employed are identical to those used for the STFT of the quasi-harmonic component. The y-axis is on a logarithmic scale. Magenta bars refer to the sum of all components; blue bars refer to the quasi-harmonic component; the yellow bars refer to the transient component, and the dark green refer to the noise component.}}
\end{figure*}

Similar behavior is found when training the model with the grand piano dataset. In the bars plot of Figure\ref{fig:grand}, the RMS mismatch is considerably lower, but the STFT one is generally slightly bigger than the upright case. This may be due to the physics equations used for the partials' decay envelope being intended for the grand piano case. The upright piano type could involve slightly different behaviors due to the different soundboards. In this case, there are no specific keys that appear to be more challenging for accurate modeling, unlike the situation with the upright piano. 

\begin{figure*}[ht!]
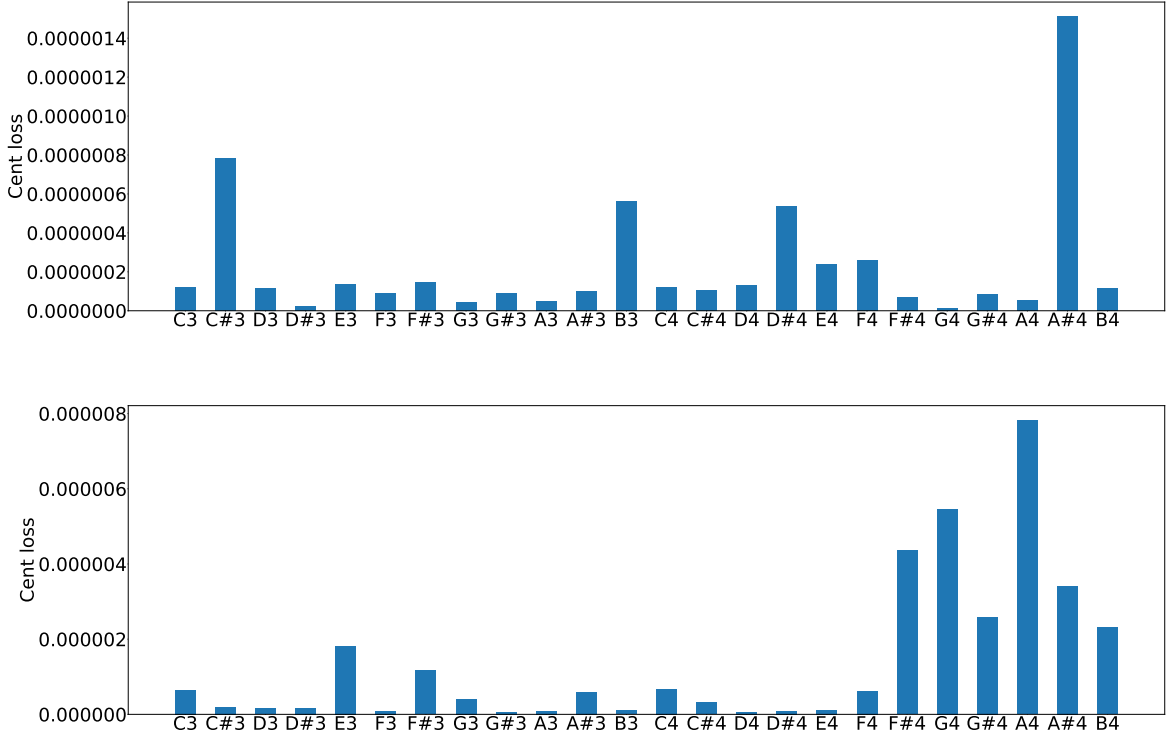

\centering
\includegraphics[scale=0.07]{Figs/cent_up.pdf}\\
\includegraphics[scale=0.07]{Figs/cent_grand.pdf}
\caption{\label{fig:cent}{\it Cent loss for all the keys of the upright piano (top) and grand piano (bottom).}}
\end{figure*}

The Cent losses, as reported in Figure~\ref{fig:cent}, indicates a good match for inharmonicity for the quasi-harmonic module, reporting a maximum cent deviation of approximately $1.81 \cdot 10^{-3}$ and $9,38 \cdot 10^{-3}$, for the upright and grand piano, respectively. However, the model starts already from a realistic estimation of the inharmonic factor, which facilitates the partial distributions. In addition, this takes into account the first $6$ partials. The higher keys are more challenging to be accurately modeled, especially for the grand piano dataset.

\begin{figure}[ht!]
\centering
\includegraphics[scale=0.33]{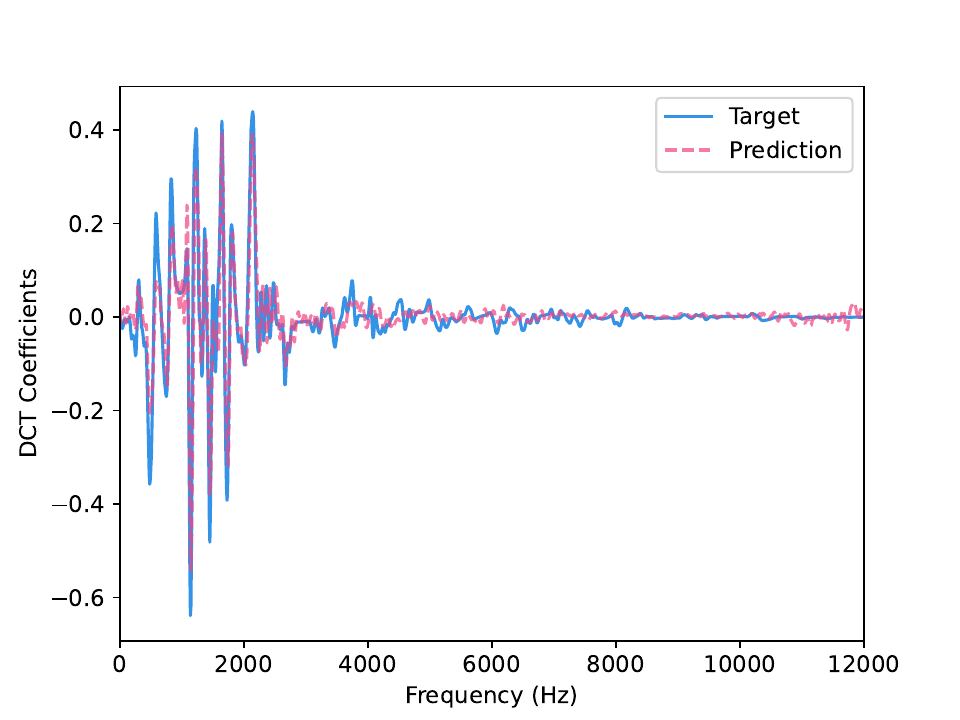}
\includegraphics[scale=0.33]{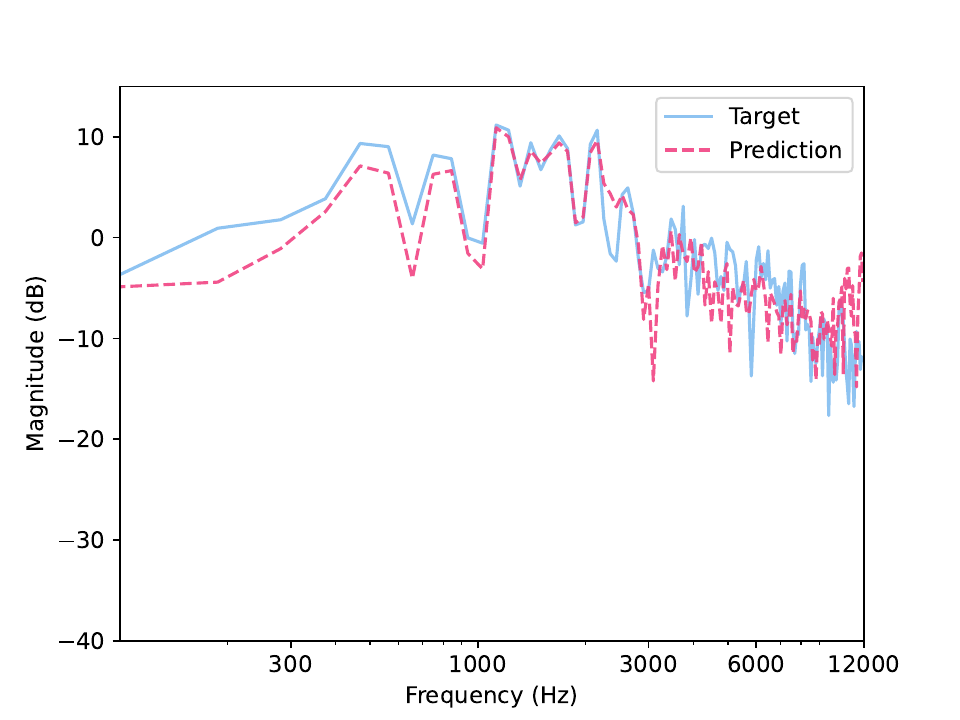}
\includegraphics[scale=0.33]{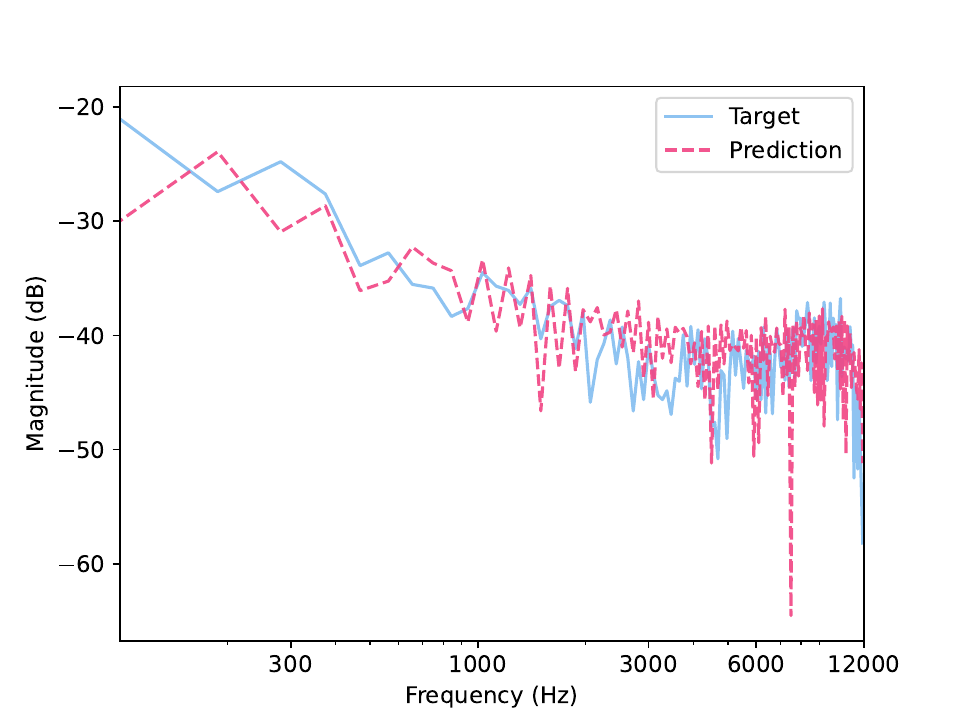}\\
\caption{\label{fig:plotsT}{\it Predictions and targets for the $C4$ key at velocity $78$ in the test set. DCT transform of the transient component (left), the spectrum of the transient (middle), and the spectrum of the noise component (right). The spectra are computed using $256$ points DFT, and the DCT transform using $1300$ points.}}
\end{figure}

\begin{figure}[ht!]
\centering
\includegraphics[scale=0.5]{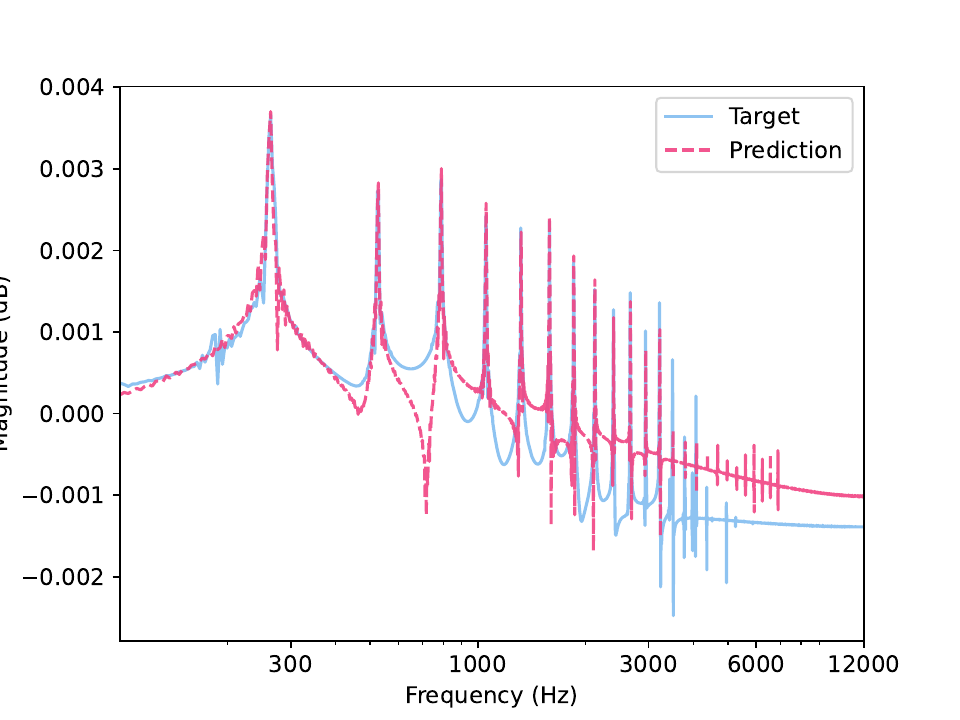}
\includegraphics[scale=0.5]{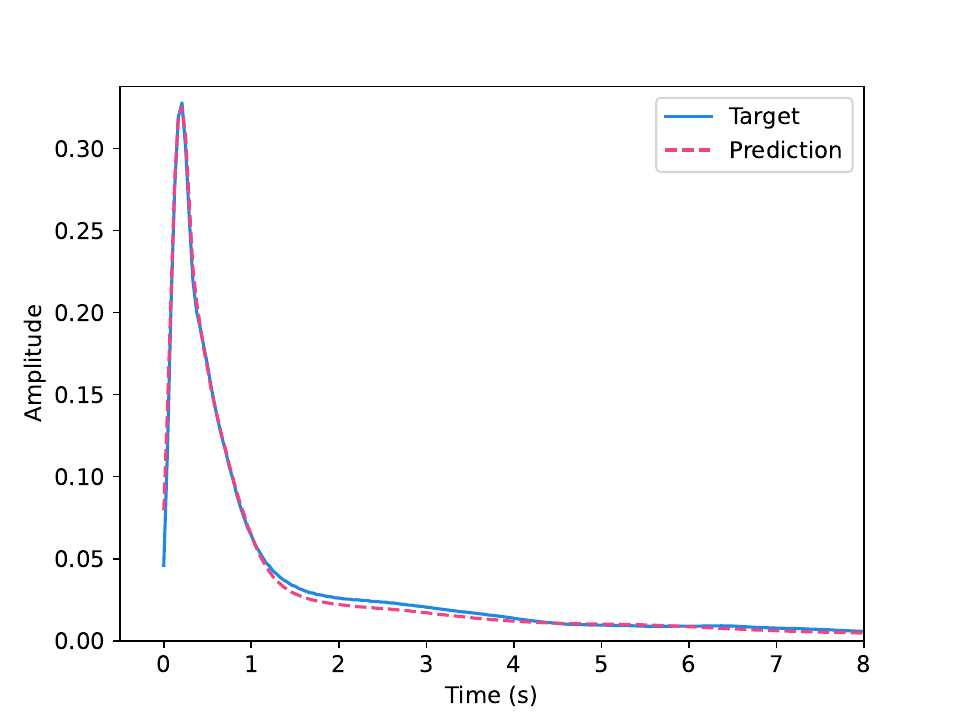}\\
\includegraphics[scale=0.5]{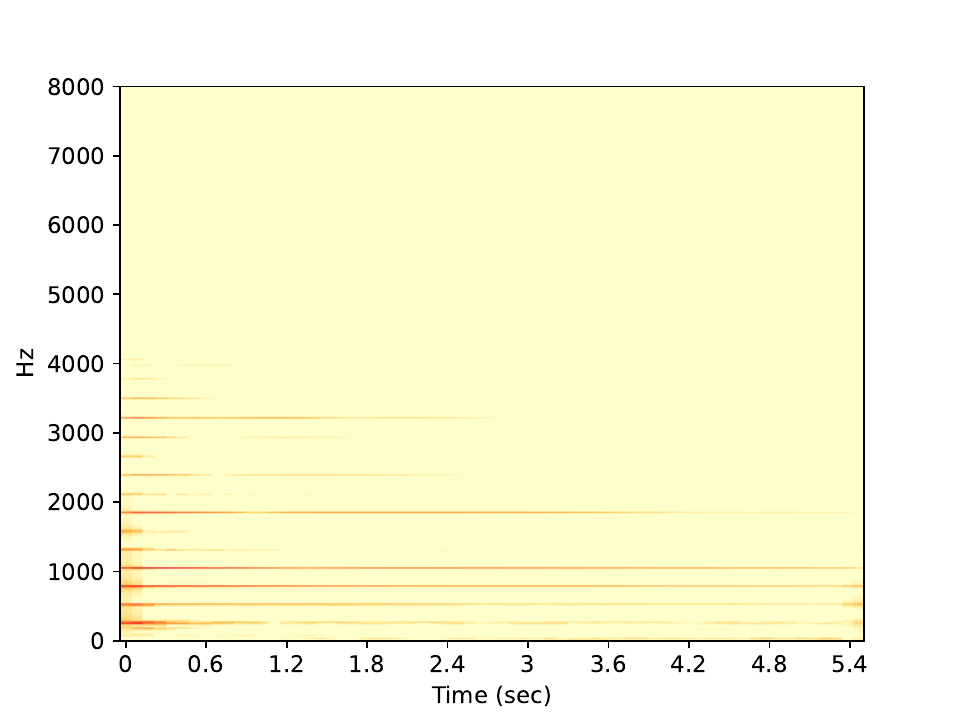}
\includegraphics[scale=0.5]{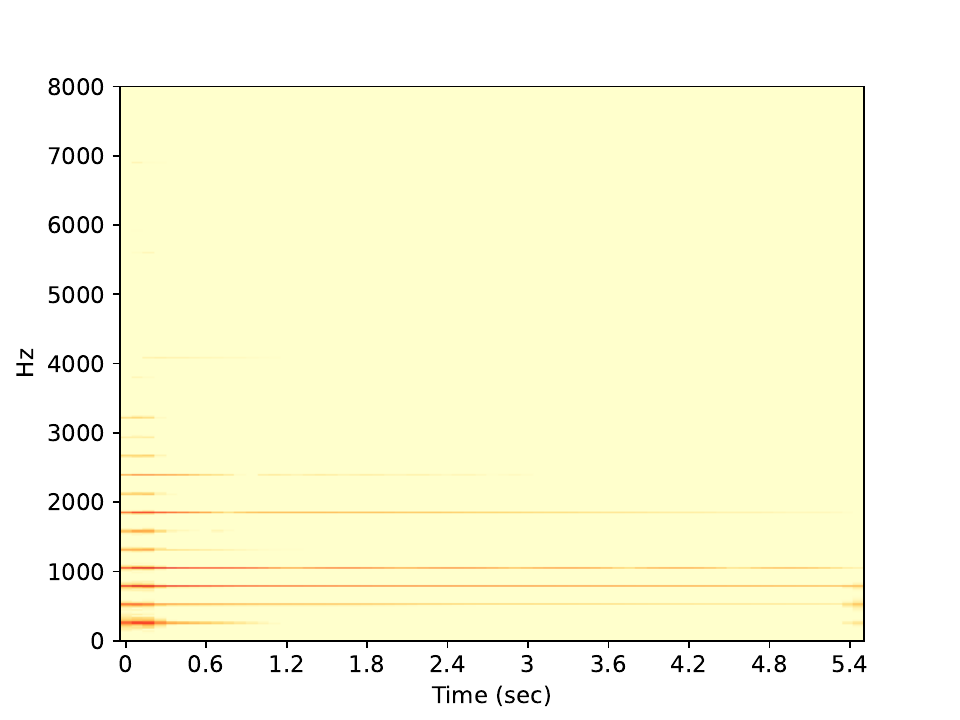}
\caption{\label{fig:plotsH}{\it Predictions and targets for the $C4$ key at velocity $78$ in the test set. Spectrum (top-left), RMS envelope (top-right), and the
spectrogram of the target (bottom-left) and prediction (bottom-right) of the sum of all components. For clarity, the frequency axis is narrowed into the range[0, $8000$] Hz. The spectra are computed using $4096$ points DFT. The RMS computation utilizes windows of $60$ samples with $25\%$ of overlap.}}
\end{figure}

Figure~\ref{fig:plotsT} shows the results for the transient and noise components of an example note in the test set, corresponding to $C4$ with velocity $78$. The predicted DCT signal is closer to the target for the lower frequencies while becoming less precise for the higher ones. This is also evident in the spectrum of the transient component, which presents a good accuracy for the low part of the spectrum, although with less energy than the target but becoming slightly less accurate at higher frequencies. On the other hand, the noise component presents a good frequency matching between the target and the prediction. 
Figure ~\ref{fig:plotsH} shows the signal's harmonic content spectrum and the RMS energy envelope. The placement of the partials in the spectra achieves an overall good accuracy even at high frequencies, in particular for the first $12$ partials, an area where precision was notably limited in our previous work ~\citep{simionato2024physics}. However, minor imprecisions at higher frequencies may still occur due to the increasing magnitude of errors in the high partials, which is attributed to the second-order term in Eq.~\ref{eq:B}. Additionally, the equation used to compute and estimate $B$ is itself an approximation.
Regarding energy, we note that the energy across the time of higher partials is less accurate, as seen in the frequency plots and spectrograms. In particular, the decay of the highest partials appears slower than that of the target. From the spectrograms, we can see that the model emulates the amplitude envelope with better accuracy in the first $10$ partials while being less accurate for the decay rate of the highest ones. The RMS profile of the target is predicted with good accuracy. In particular, the prediction shows the double decay stages, and the attack envelope also shows a good match. These aspects also represent a significant improvement over our previous work ~\citep{simionato2024physics}.

\begin{figure}[ht!]
\centering
\includegraphics[scale=0.33]{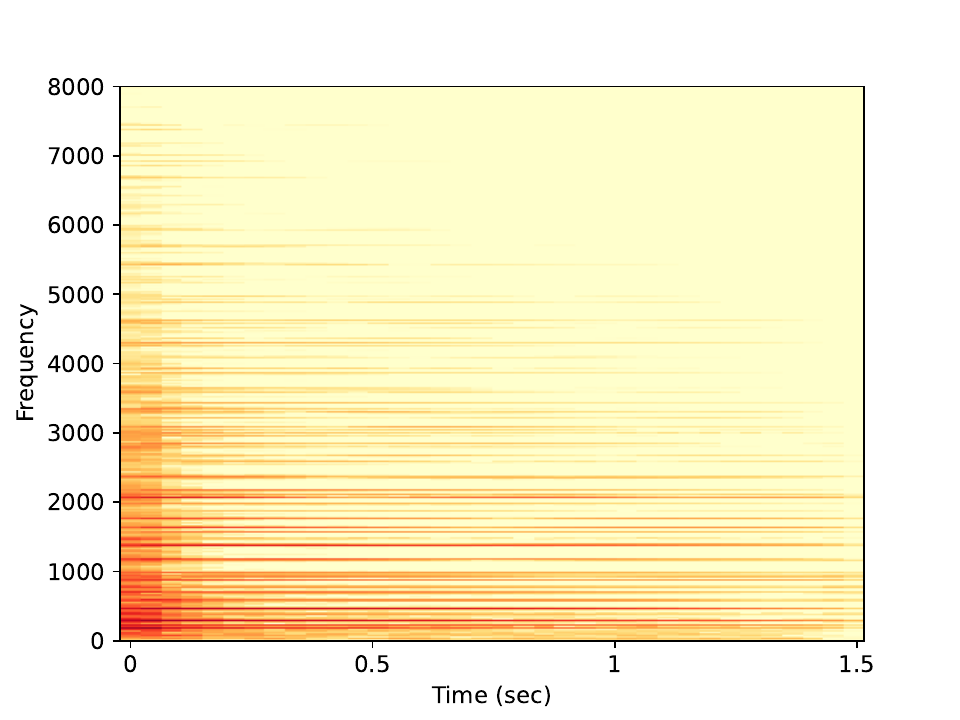}
\includegraphics[scale=0.33]{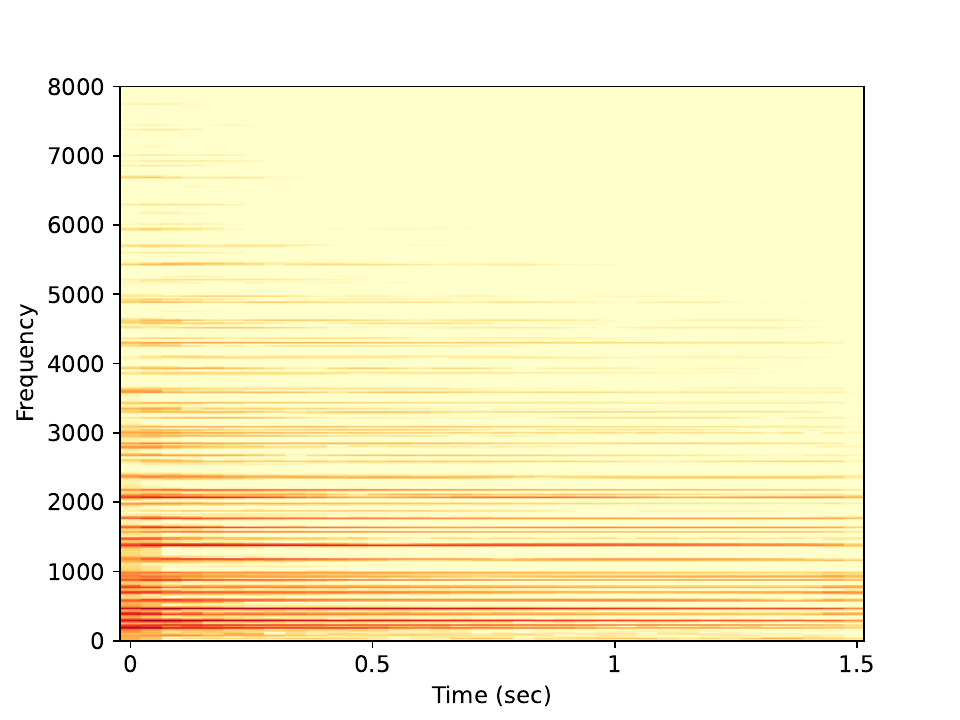}
\includegraphics[scale=0.33]{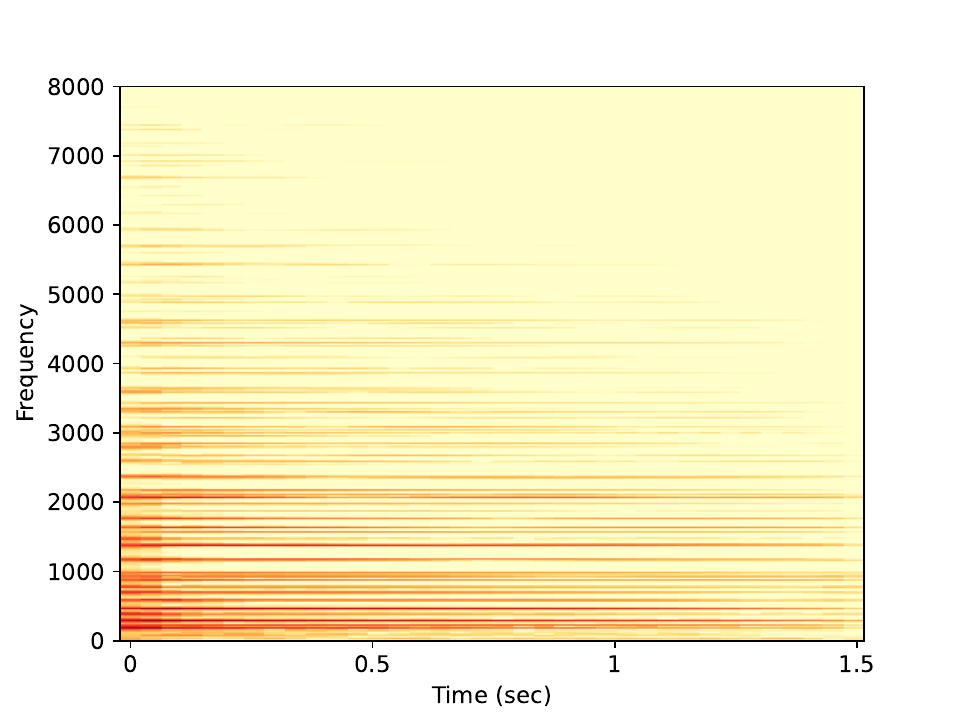}
\caption{\label{fig:chord}{\it
Spectrogram of the $G3$-$A\#3$-$D4$ trichord at velocity $80$ in the test set for the sum of single notes (left), the real recorded trichord (middle), and the model's prediction (right). The spectra are computed using $4096$ points DFT.}}
\end{figure}

Table~\ref{tab:obj} reports the losses for the trichord scenarios. The measure indicates a good similarity between the prediction and the target, which is confirmed by the spectrograms of a trichord example in the test set: the $G3$-$A\#3$-$D4$ trichord at velocity $80$. From the spectrograms shown in Figure~\ref{fig:chord}, it is evident that the simple sum presents more energy damping than the real trichord and the related prediction. In addition, the real trichord includes more frequency components and variations in the energy among the frequencies, an effect due to coupling among the three played keys. This aspect is not present in the trichord generated by the sum of notes, while the prediction of the proposed model is emulating this as well.
\begin{table}[ht]
  \caption{\itshape Test set losses for the trichord dataset.}
\centering
\begin{tabular}{|c|c|c|}
\hline
Module & Type & Loss\\\hline
\hline
Trichords  & STFT loss  & $8.32 \cdot 10^{-1}$\\
 & RMS loss  & $ 1.63 \cdot 10^{-3}$\\
 \hline
\end{tabular}
\label{tab:obj}
\end{table}

\begin{figure*}[ht]
\centering
\includegraphics[scale=0.45]{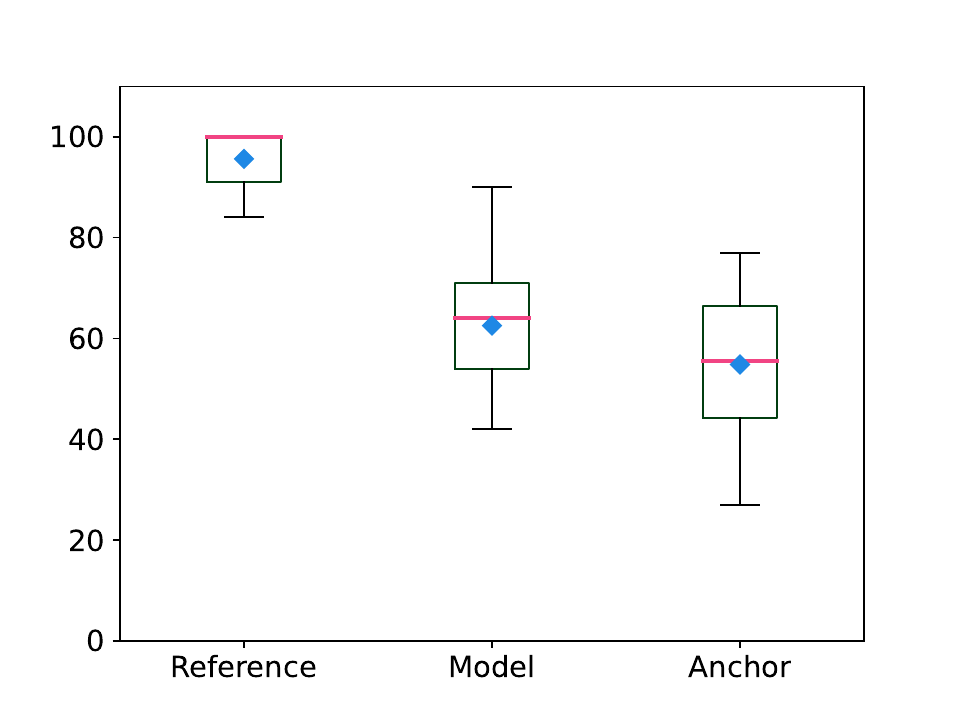}
\includegraphics[scale=0.45]{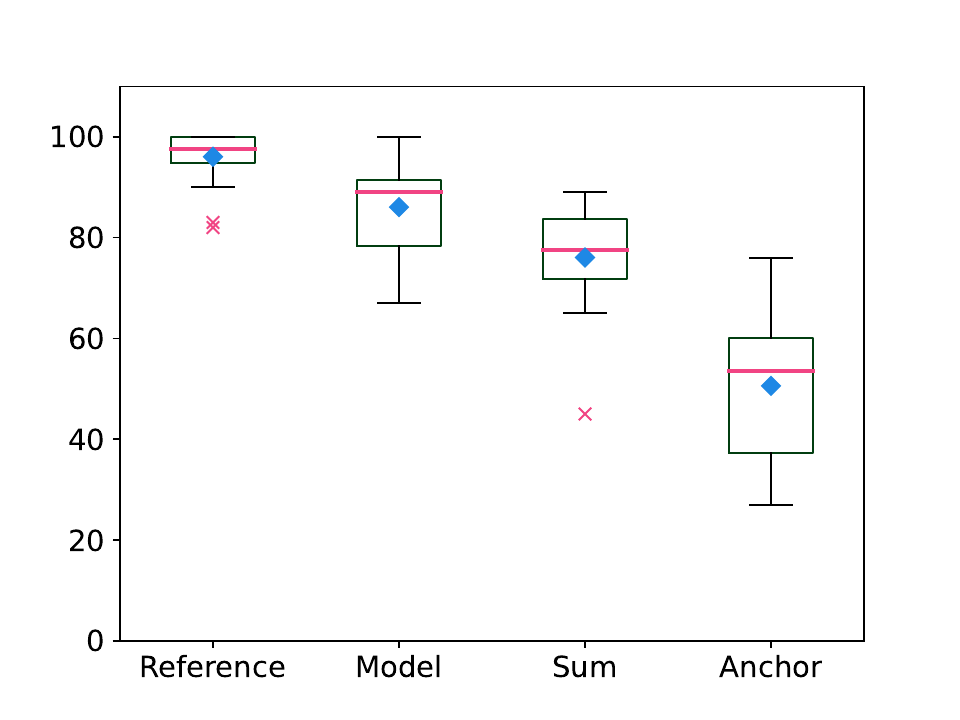}
\caption{\label{fig:lis}{\it Box plot showing the rating results of the listening tests for single notes (left) and trichords (right). The magenta horizontal lines and the diamond markers indicate the median and the mean values, respectively. Outliers are displayed as crosses. }}
\end{figure*}

A total of $20$ participants took part in the MUSHRA test, consisting of $13$ men and $7$ women, aged between $26$ and $36$. All participants were trained musicians with varying levels of expertise, ranging from intermediate to expert. None reported any hearing impairment, and none met the criteria for exclusion due to clear and obvious rating errors. Figure ~\ref{fig:lis} presents the results of the test, where the box plots illustrate the median, lower, and upper quartiles, as well as the maximum, minimum values, and outliers of the responses. It is evident that the sound generated by the model is perceptually distinguishable from that of a real piano, which was used to train the model, but the difference is not substantial. For the chords, the model provides a better sound than the simple sum of the individual notes, typical of most synthesis approaches. Table ~\ref{tab:stats} presents a statistical analysis of the results based on the non-parametric Friedman test and Cliff's delta. The results show small $p$-values below the statistical significance threshold of $0.05$ and predominantly large effect sizes for the listening test results on both single notes and trichords, indicating statistical significance and practical relevance.

\begin{table}[ht]
  \caption{\itshape $p$-value and Cliff'delta $|d|$ for the MUSHRA test on single notes and trichords.}
\centering
\begin{tabular}{|c|c|c|}
\hline
Case & $p$-value & $|d|$\\\hline
\hline
\textit{Single Notes} & $1.04 \cdot 10^{-8}$ & \\
\hline
reference - model & &  1.00 (large)\\
reference - anchor & &  1.00 (large)\\
model - anchor & &  0.25 (small)\\
\hline
\hline
\textit{Trichords}  & $6.26 \cdot 10^{-14}$ & \\
\hline
reference - model & &  0.92 (large)\\
reference - anchor & &  1.00 (large)\\
model - anchor & &  1.00 (large)\\
sum - anchor & &  1.00 (large)\\
sum - reference  & &  0.98 (large)\\
sum - model & &  0.42 (medium)\\
model - anchor & &  0.94 (large)\\
 \hline
\end{tabular}
\label{tab:stats}
\end{table}

\begin{figure}[ht!]
\centering
\includegraphics[scale=0.45]{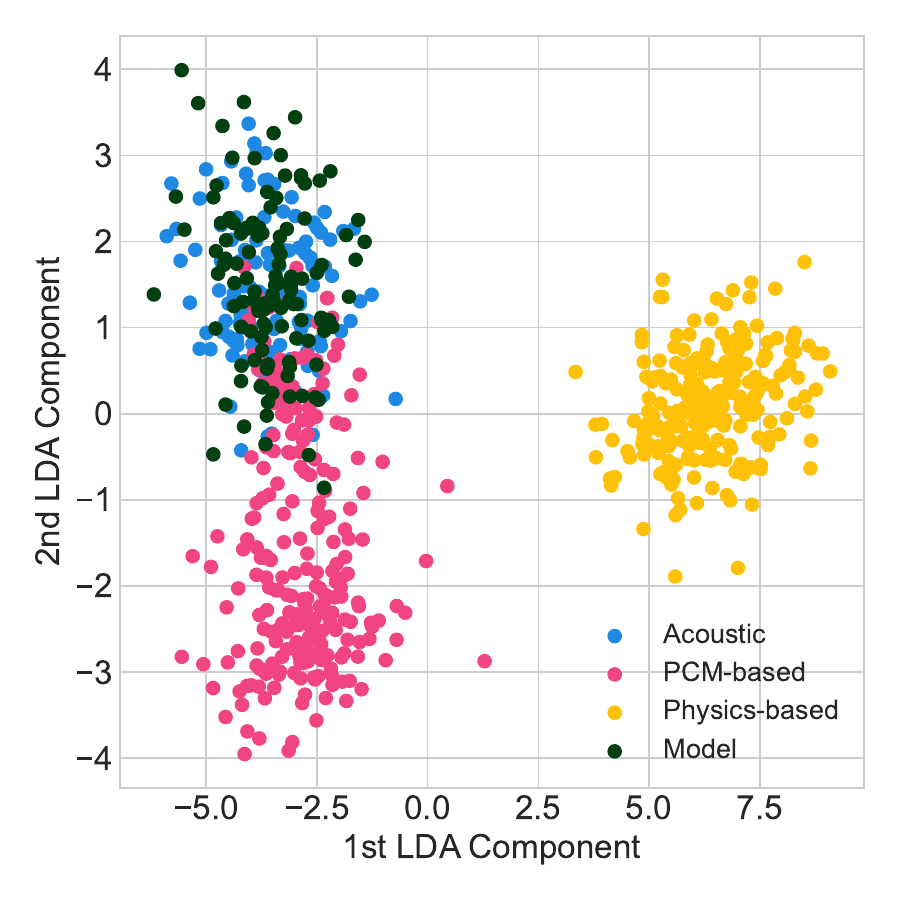}
\includegraphics[scale=0.45]{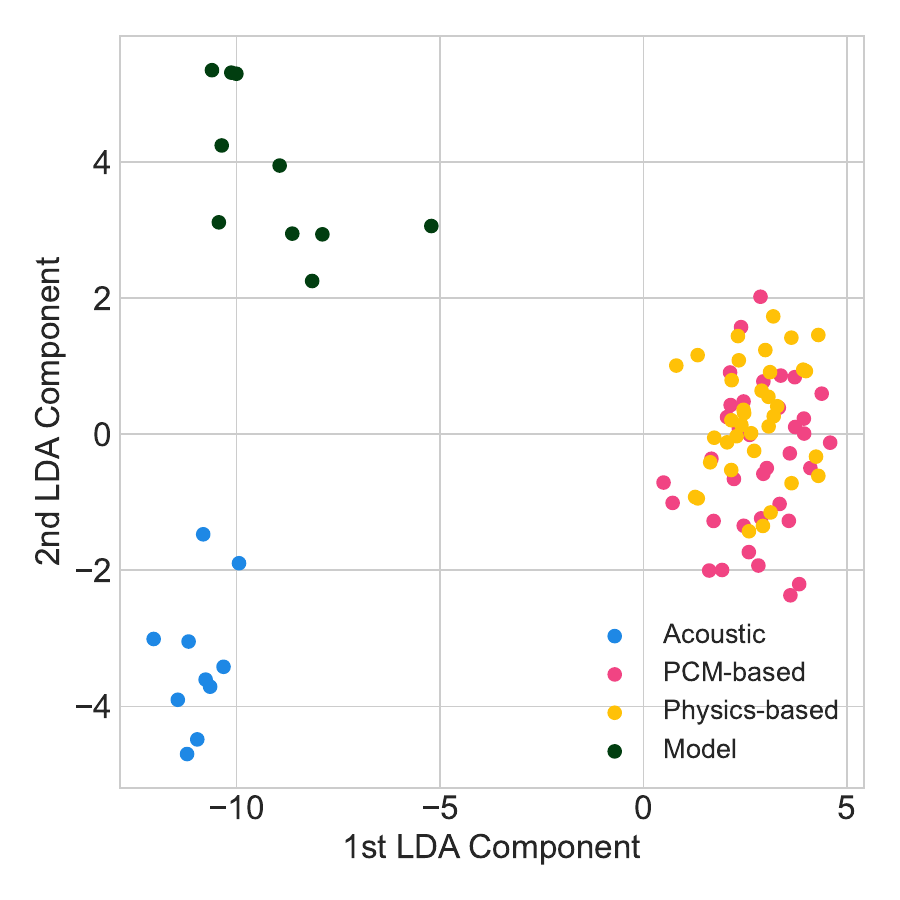}
\caption{\label{fig:comparative}{\it
Projected feature vectors to 2D space using LDA
for the case of single-note stimuli (left) and trichords (right).}}
\end{figure}

Figure \ref{fig:comparative} reports the related LDA projections for the single notes and trichords. In the first case, the comparison focuses on the sound variation at different velocities, while the second considers the difference between the sum and the actual trichord. The examples used from the models are included in the test set. We can see how, for single notes, the models' predictions are confined in the same cluster of the acoustic piano, while the samples-based and physics-based digital piano examples are in different visible clusters. This suggests that the proposed methodology is able to simulate the sound changes due to different velocities. 
On the other hand, the trichords case shows that the acoustic cluster is separated from the model's one, although the model cluster does not overlap with samples-based and physics-based clusters, which in this case are totally overlapped. Therefore, although the model behaves differently from the samples-based and physics models, whose emulations are close to the simple sum, it presents differences in the frequency components with respect to the acoustic case.

\subsection{Efficiency}

Generating the damping parameters based on physical laws helps to restrict the layer sizes and, consequently, reduces the computational cost. Furthermore, utilizing target information such as the inharmonicity factor to initialize the training simplifies the learning challenges for the model, further contributing to the achievement of accuracy with smaller neural networks. The Quasi-Harmonic module comprises only $7,601$ trainable parameters and requires $\approx 18,272$ FLOPs/sample. In contrast, the Transient module utilizes several convolutional layers to achieve good accuracy. It has a total of $5,589$ trainable parameters and requires $\approx 40,862 $ FLOPs per waveform ($\approx 31$ FLOPs/sample). Lastly, the noise module presents $484$ trainable parameters and $\approx 728$ FLOPs/sample. Regarding memory, considering the single-precision floating-point format, the network parameters require $\approx 60$ megabytes. Compared to~\citep{renault2022differentiable}, our model has $127.6$k against $281.5$k trainable parameters (excluding those related to the reverb) considering a polyphony level of $16$.

\section{Discussion \& Conclusions}
\label{sec:conclusion}

The paper presented a novel method to approach differentiable spectral modeling. The target sound, in this case, piano notes, is decomposed into harmonic, transient, and noise components. The harmonic part is synthesized using sinusoidal modeling and physics knowledge. From MIDI information, the trained model predicts the inharmonicity factor, the frequency-dependent damping coefficients, and the attack time. The inharmonicity factor and the frequency-dependent damping coefficients are computed using physical equations. A sine generator computes the partials multiplied by the amplitude envelope to generate the quasi-harmonic content. Beatings, double decay, and even phantom partials are included in the model, generating other sets of partials. The noise is targeted by predicting a time-varying filter to filter the white noise, its mean, and amplitude envelope across time. Finally, the transient component is generated using a deep convolutional oscillator that produces the discrete cosine transform (DCT) of the transient, which is later reversed. The network generating the DCT consists of a stack of upsampling and convolutional layers that generate a waveform from the MIDI information. 
The model is trained to emulate a specific piano by learning directly from actual sound recordings and analytical information extracted from them. Eventual imprecision in the analysis algorithms, physics-derived equations, or unconsidered physical aspects can negatively impact the resulting model's accuracy. For example, at this stage, the soundboard influence is implicitly learned from the damping layer, and the room reverb is not included. The model learns three components separately, lowering the task's complexity. The physical law guiding the generation of the harmonic content also helps lower the computational cost requirements. The architecture is designed to generate an arbitrary number of samples per inference iteration, thereby allowing flexibility and minimizing input-to-output latency. For this specific configuration, we have demonstrated the integration of more complex aspects, such as the coupling effect when different keys are struck simultaneously in trichords.

All three sub-models can provide overall good emulation accuracy. The quasi-harmonic content matches the partial placements, although it becomes less precise at higher partial. Frequently, these are predicted to have a greater energy than the target piano recordings. The RMS envelopes show a good matching as well, and the predicted signal presents two decay rates. Also, the generation of the transient and noisy parts presents overall spectral accuracy. On the other hand, listening tests still underscore perceptual differences between the target and emulated piano sounds. In particular, the attack portion of the note is where most inaccuracies appear to be concentrated. This is evident when listening to the audio examples we have made publicly available, and it was also highlighted during informal post-test discussions with the majority of participants.

The computational bottleneck of the model is represented by the transient generation that requires a relatively large neural network. The discrete cosine transform allows the restriction of the size of the vector to be generated since the limited extension of the frequency content and, in turn, lowers the network's effort. Finally, the method relies on the harmonic, transient, and noise components separation quality, which is crucial for learning, and on the precision of the utilized physical laws. This aspect introduces further flexibility to the model. Once trained, the model accepts variations on its equation, and users can potentially navigate different equations, such as, in this case, the attack envelope and energy damping. The possibility of emulating the sound of individual piano notes without the requirements of storing audio samples can also be a memory-efficient alternative to the sample-based synthesizer. In the future, we will develop datasets to integrate other complex performance scenarios, such as re-stricken keys, arpeggios, and chords consisting of different numbers of notes and arbitrarily simultaneously played notes. Another aspect to further investigate is the inclusion of phase information in the quasi-harmonic model, which could potentially enhance the accuracy of note attack. In the current model, convergence issues arise and gradients tend to explode when phase information is incorporated—that is, when both the frequency and phase of each partial are predicted. The underlying reasons for this phenomenon remain unknown; however, they may lie in the cyclic nature of the phase.

\bibliographystyle{authordate1}
\bibliography{references}  

\begin{thebibliography}{}

\bibitem[\protect\citename{Bank \& Sujbert, }2003]{bank2003modeling}
Bank, Bal{\'a}zs, \& Sujbert, L{\'a}szl{\'o}. 2003.
\newblock Modeling the longitudinal vibration of piano strings.
\newblock {\em Pages  143--146 of:} {\em Proc. Stockholm Music Acoust. Conf}.

\bibitem[\protect\citename{Bank \& Sujbert, }2005]{bank2005generation}
Bank, Bal{\'a}zs, \& Sujbert, L{\'a}szl{\'o}. 2005.
\newblock Generation of longitudinal vibrations in piano strings: From physics to sound synthesis.
\newblock {\em The Journal of the Acoustical Society of America}, {\bf 117}(4), 2268--2278.

\bibitem[\protect\citename{Birnbaum {\em et~al.}, }2019]{birnbaum2019temporal}
Birnbaum, S., Kuleshov, V., Enam, Z., Koh, P. W.~W, \& Ermon, S. 2019.
\newblock Temporal FiLM: Capturing Long-Range Sequence Dependencies with Feature-Wise Modulations.
\newblock {\em Advances in Neural Information Processing Systems}, {\bf 32}.

\bibitem[\protect\citename{Caspe {\em et~al.}, }2022]{caspe2022ddx7}
Caspe, F., McPherson, A., \& Sandler, M. 2022.
\newblock Ddx7: Differentiable fm synthesis of musical instrument sounds.
\newblock {\em arXiv preprint arXiv:2208.06169}.

\bibitem[\protect\citename{Chabassier {\em et~al.}, }2014]{chabassier2014time}
Chabassier, J., Chaigne, A., \& Joly, P. 2014.
\newblock Time domain simulation of a piano. Part 1: model description.
\newblock {\em ESAIM: Mathematical Modelling and Numerical Analysis}, {\bf 48}(5), 1241--1278.

\bibitem[\protect\citename{Conklin~Jr, }1999]{conklin1999generation}
Conklin~Jr, Harold~A. 1999.
\newblock Generation of partials due to nonlinear mixing in a stringed instrument.
\newblock {\em The Journal of the Acoustical Society of America}, {\bf 105}(1), 536--545.

\bibitem[\protect\citename{Cuesta \& Valette, }1988]{valette1988evolution}
Cuesta, C, \& Valette, C. 1988.
\newblock Evolution temporelle de la vibration des cordes de clavecin.
\newblock {\em Acta Acustica united with Acustica}, {\bf 66}(1), 37--45.

\bibitem[\protect\citename{Dauphin {\em et~al.}, }2017]{dauphin2017language}
Dauphin, Yann~N, Fan, Angela, Auli, Michael, \& Grangier, David. 2017.
\newblock Language modeling with gated convolutional networks.
\newblock {\em Pages  933--941 of:} {\em International conference on machine learning}.
\newblock PMLR, Sydney, Australia.

\bibitem[\protect\citename{Driedger {\em et~al.}, }2014]{driedger2014extending}
Driedger, J., M{\"u}ller, M., \& Disch, S. 2014.
\newblock Extending Harmonic-Percussive Separation of Audio Signals.
\newblock {\em In:} {\em ISMIR}.

\bibitem[\protect\citename{Ege, }2009]{ege2009table}
Ege, Kerem. 2009.
\newblock {\em La table d'harmonie du piano-{\'E}tudes modales en basses et moyennes fr{\'e}quences}.
\newblock Ph.D. thesis, Ecole Polytechnique X, Paris, France.

\bibitem[\protect\citename{Engel {\em et~al.}, }2020]{engel2020ddsp}
Engel, J., Hantrakul, L., Gu, C., \& Roberts, A. 2020.
\newblock DDSP: Differentiable digital signal processing.
\newblock {\em Int. Conf. on Learning Representations}.

\bibitem[\protect\citename{Etchenique {\em et~al.}, }2015]{etchenique2015coupling}
Etchenique, N., Collin, S.~R, \& Moore, T.~R. 2015.
\newblock Coupling of transverse and longitudinal waves in piano strings.
\newblock {\em J. Acoust. Soc. Am.}, {\bf 137}(4), 1766--1771.

\bibitem[\protect\citename{Hawthorne {\em et~al.}, }2019]{hawthorne2018enabling}
Hawthorne, C., Stasyuk, A., Roberts, A., Simon, I., Huang, C.~A., Dieleman, S., Elsen, E., {\em et~al.} 2019.
\newblock Enabling factorized piano music modeling and generation with the MAESTRO dataset.
\newblock {\em Int. Conf. on Learning Representations}.

\bibitem[\protect\citename{Hendrycks \& Gimpel, }2023]{hendrycks2023gaussian}
Hendrycks, D., \& Gimpel, K. 2023.
\newblock Gaussian Error Linear Units (GELUs).

\bibitem[\protect\citename{Issanchou {\em et~al.}, }2017]{issanchou2017modal}
Issanchou, C., Bilbao, S., Le~Carrou, JL., Touz{\'e}, C., \& Doar{\'e}, O. 2017.
\newblock A modal-based approach to the nonlinear vibration of strings against a unilateral obstacle: Simulations and experiments in the pointwise case.
\newblock {\em J. Sound Vibr.}, {\bf 393}, 229--251.

\bibitem[\protect\citename{ITU-R, }2015]{bs20151534}
ITU-R. 2015.
\newblock BS.1534-3, Method for the subjective assessment of intermediate quality level of audio systems.
\newblock {\em International Telecommunication Union, Geneva, Switzerland}, Oct.

\bibitem[\protect\citename{Jonason {\em et~al.}, }2023]{jonason2023ddsp}
Jonason, N., Wang, X., Cooper, E., Juvela, L., Sturm, B.~LT, \& Yamagishi, J. 2023.
\newblock DDSP-based Neural Waveform Synthesis of Polyphonic Guitar Performance from String-wise MIDI Input.
\newblock {\em arXiv preprint arXiv:2309.07658}.

\bibitem[\protect\citename{Kawamura {\em et~al.}, }2022]{kawamura2022differentiable}
Kawamura, M., Nakamura, T., Kitamura, D., Saruwatari, H., Takahashi, Y., \& Kondo, K. 2022.
\newblock Differentiable digital signal processing mixture model for synthesis parameter extraction from mixture of harmonic sounds.
\newblock {\em In:} {\em IEEE Int. Conf. on Acoustics, Speech and Signal Processing (ICASSP)}.

\bibitem[\protect\citename{Kingma \& Ba, }2014]{kingma2014adam}
Kingma, D.~P., \& Ba, J. 2014.
\newblock Adam: A method for stochastic optimization.
\newblock {\em Int. Conf. on Learning Representations}.

\bibitem[\protect\citename{Krekovi{\'c}, }2022]{krekovic2022deep}
Krekovi{\'c}, G. 2022.
\newblock Deep Convolutional Oscillator: Synthesizing Waveforms from Timbral Descriptors.
\newblock {\em Pages  200--206 of:} {\em Proceedings of the International Conference on Sound and Music Computing (SMC)}.

\bibitem[\protect\citename{Liu {\em et~al.}, }2023]{liu2023ddsp}
Liu, Y., Jin, C., \& Gunawan, D. 2023.
\newblock DDSP-SFX: Acoustically-guided sound effects generation with differentiable digital signal processing.
\newblock {\em In:} {\em Proc. of the Conf. on Digital Audio Effects (DAFx)}.

\bibitem[\protect\citename{Mezza {\em et~al.}, }2024]{mezza2024data}
Mezza, Alessandro~Ilic, Giampiccolo, Riccardo, De~Sena, Enzo, \& Bernardini, Alberto. 2024.
\newblock Data-driven room acoustic modeling via differentiable feedback delay networks with learnable delay lines.
\newblock {\em EURASIP Journal on Audio, Speech, and Music Processing}, {\bf 2024}(1), 51.

\bibitem[\protect\citename{Nakamura, }1994]{nakamura1994characteristics}
Nakamura, I. 1994.
\newblock Characteristics of piano sound spectra.
\newblock {\em Proc. SMAC93}.

\bibitem[\protect\citename{Pascanu {\em et~al.}, }2013]{pascanu2013difficulty}
Pascanu, R., Mikolov, T., \& Bengio, Y. 2013.
\newblock On the difficulty of training recurrent neural networks.
\newblock {\em In:} {\em Int. Conf. on machine learning}.

\bibitem[\protect\citename{Podlesak \& Lee, }1988]{podlesak1988dispersion}
Podlesak, M., \& Lee, A.~R. 1988.
\newblock Dispersion of waves in piano strings.
\newblock {\em J. Acoust. Soc. Am.}, {\bf 83}(1), 305--317.

\bibitem[\protect\citename{Renault {\em et~al.}, }2022]{renault2022differentiable}
Renault, L., Mignot, R., \& Roebel, A. 2022.
\newblock Differentiable Piano Model for MIDI-to-Audio Performance Synthesis.
\newblock {\em In:} {\em Proc. of the Conf. on Digital Audio Effects (DAFx)}.

\bibitem[\protect\citename{Schoeffler {\em et~al.}, }2018]{schoeffler2018webmushra}
Schoeffler, M., Bartoschek, S., St{\"o}ter, FR, Roess, M., Westphal, S., Edler, B., \& Herre, J. 2018.
\newblock webMUSHRA—A comprehensive framework for web-based listening tests.
\newblock {\em Journal of Open Research Software}, {\bf 6}(1), 8.

\bibitem[\protect\citename{Serra \& Smith, }1990]{serra1990spectral}
Serra, X., \& Smith, J. 1990.
\newblock Spectral modeling synthesis: A sound analysis/synthesis system based on a deterministic plus stochastic decomposition.
\newblock {\em Computer Music Journal}, {\bf 14}(4), 12--24.

\bibitem[\protect\citename{Shan {\em et~al.}, }2022]{shan2022differentiable}
Shan, S., Hantrakul, L., Chen, J., Avent, M., \& Trevelyan, D. 2022.
\newblock Differentiable wavetable synthesis.
\newblock {\em In:} {\em IEEE Int. Conf. on Acoustics, Speech and Signal Processing (ICASSP)}.

\bibitem[\protect\citename{Shier {\em et~al.}, }2023]{shier2023differentiable}
Shier, J., Caspe, F., Robertson, A., Sandler, M., Saitis, C., \& McPherson, A. 2023.
\newblock Differentiable modelling of percussive audio with transient and spectral synthesis.
\newblock {\em Forum Acusticum}.

\bibitem[\protect\citename{Simionato \& Fasciani, }2023]{simionato2023computational}
Simionato, R., \& Fasciani, S. 2023.
\newblock A Comparative Computational Approach to Piano Modeling Analysis.
\newblock {\em In:} {\em Proceedings of the International Conference on Sound and Music Computing (SMC)}.

\bibitem[\protect\citename{Simionato \& Fasciani, }2024]{simionato2024conditioning}
Simionato, R., \& Fasciani, S. 2024.
\newblock Conditioning Methods for Neural Audio Effects.
\newblock {\em In:} {\em Proceedings of the International Conference on Sound and Music Computing (SMC)}.

\bibitem[\protect\citename{Simionato {\em et~al.}, }2024]{simionato2024physics}
Simionato, R., Fasciani, S., \& Holm, S. 2024.
\newblock Physics-informed differentiable method for piano modeling.
\newblock {\em Frontiers in Signal Processing}, {\bf 3}, 1276748.

\bibitem[\protect\citename{Tan, }2017]{tan2017piano}
Tan, J.~J. 2017.
\newblock {\em Piano acoustics: string’s double polarisation and piano source identification}.
\newblock Ph.D. thesis, Universit{\'e} Paris Saclay (COmUE), Paris, France.

\bibitem[\protect\citename{Verma \& Meng, }2000]{verma2000extending}
Verma, T.~S, \& Meng, T.~HY. 2000.
\newblock Extending spectral modeling synthesis with transient modeling synthesis.
\newblock {\em Computer Music Journal},  47--59.

\bibitem[\protect\citename{Weinreich, }1977]{weinreich1977coupled}
Weinreich, G. 1977.
\newblock Coupled piano strings.
\newblock {\em J. Acoust. Soc. Am.}, {\bf 62}(6), 1474--1484.

\bibitem[\protect\citename{Wiggins \& Kim, }2023]{wiggins2023differentiable}
Wiggins, A., \& Kim, Y. 2023.
\newblock A Differentiable Acoustic Guitar Model for String-Specific Polyphonic Synthesis.
\newblock {\em In:} {\em 2023 IEEE Workshop on Applications of Signal Processing to Audio and Acoustics (WASPAA)}.

\end{thebibliography}

\end{document}